\documentclass[a4paper]{spie}  
\usepackage[]{graphicx}
\usepackage{amsmath,xspace}
\usepackage{multirow}
\usepackage{array}

\DeclareTextSymbol{\degre}{OT1}{23}
\newcommand{\mum}{\mbox{{\usefont{U}{eur}{m}{n}{\char22}}m}\xspace}

\title{PERSEE: Experimental results on the cophased nulling bench}

\author{J. Lozi\supit{a,b}, F. Cassaing\supit{a,b}, J.M. Le Duigou\supit{c}, K. Houairi\supit{a,b}, B. Sorrente\supit{a,b}, J. Montri\supit{a,b}, S. Jacquinod\supit{d}, J-M Reess\supit{e,b}, L. Pham\supit{e,b}, E. Lhom\'e\supit{e,b}, T. Buey\supit{e,b}, F. H\'enault\supit{f}, A. Marcotto\supit{f}, P. Girard\supit{f}, N. Mauclert\supit{f}, M. Barillot\supit{g}, V. Coud\'e du Foresto\supit{e,b} and M. Ollivier\supit{d}
\skiplinehalf
\supit{a}ONERA, BP72 - 29 avenue de la Division Leclerc, 92322 Ch\^atillon Cedex, France;\\
\supit{b}Groupement d'Int\'er\^et Scientifique PHASE (Partenariat Haute r\'esolution Angulaire Sol Espace) between ONERA, Observatoire de Paris, CNRS and Universit\'e Paris Diderot;\\
\supit{c}CNES - Centre National d'Etudes Spatiales, Toulouse, France;\\
\supit{d}IAS - Institut d'Astrophysique Spatiale, Orsay, France;\\
\supit{e}Observatoire de Paris - LESIA, Meudon, France;\\
\supit{f}Observatoire de la C\^ote d'Azur, Grasse, France;\\
\supit{g}Thal\`es Al\'enia Space, Cannes, France
}

\authorinfo{E-mail: julien.lozi@onera.fr, Telephone: +33 (0)1 46 73 47 59}

\begin{document}
  \maketitle

\begin{abstract}
Nulling interferometry is still a promising method to characterize spectra of exoplanets. One of the main issues is to cophase at a nanometric level each arm despite satellite disturbances. The bench PERSEE aims to prove the feasibility of that technique for spaceborne missions. After a short description of PERSEE, we will first present the results obtained in a simplified configuration: we have cophased down to 0.22~nm~rms in optical path difference (OPD) and 60~mas~rms in tip/tilt, and have obtained a monochromatic null of $3\cdot10^{-5}$ stabilized at $3\cdot10^{-6}$. The goal of 1~nm with additional typical satellite disturbances requires the use of an optimal control law; that is why we elaborated a dedicated Kalman filter. Simulations and experiments show a good rejection of disturbances. Performance of the bench should be enhanced by using a Kalman control law, and we should be able to reach the desired nanometric stability. Following, we will present the first results of the final polychromatic configuration, which includes an achromatic phase shifter, perturbators and optical delay lines. As a conclusion, we give the first more general lessons we have already learned from this experiment, both at system and component levels for a future space mission.
\end{abstract}



\section{INTRODUCTION}
\label{sec:intro}  

Although it has been recently postponed due to high cost and risks, nulling interferometry in space remains one of the very few direct detection methods able to characterize extrasolar planets and particularly telluric ones. Within this framework, several projects such as DARWIN\cite{Leger96,Leger07}, TPF-I\cite{Beichman99,Lawson00}, FKSI\cite{Danchi06} or PEGASE\cite{Ollivier07,Leduigou06b} have been proposed in the past years. Most of them are based on a free flying concept. It allows firstly to avoid atmosphere turbulence, and secondly to distribute instrumental function over many satellites flying in close formation. In this way, a very high angular resolution can be achieved with an acceptable launch mass. But the price to pay is to very precisely position and stabilize relatively the spacecrafts, in order to achieve a deep and stable extinction of the star. Understanding and mastering all these requirements are great challenges and key issues towards the feasibility of these missions. Thus, we decided to experimentally study this question and focus on some possible simplifications of the concept.

Since 2006, PERSEE (PEGASE Experiment for Research and Stabilization of Extreme Extinction)\cite{Cassaing08}laboratory test bench is under development by a consortium composed of Centre National d'Etudes Spatiales (CNES), Institut d'Astrophysique Spatiale (IAS), Observatoire de Paris-Meudon (LESIA), Observatoire de la C\^ote d'Azur (OCA), Office National d'Etudes et de Recherches A\'erospatiales (ONERA), and Thal\`es Al\'enia Space (TAS). It is mainly funded by CNES R\&D. PERSEE couples an infrared wide band nulling interferometer with local OPD and tip/tilt control loops and a free flying Guidance Navigation and Control (GNC) simulator able to introduce realistic disturbances. Although it was designed in the framework of the PEGASE free flying space mission, PERSEE can adapt very easily to other contexts like FKSI (in space, with a 10 m long beam structure) or ALADDIN\cite{Coudeduforesto06} (on ground, in Antarctica) because the optical designs of all those missions are very similar.

After a short description of the experimental setup, we will present first the results obtained in an intermediate configuration with monochromatic light. Then, the key issue of filtering laboratory disturbances will be described in some details before presenting our very first and preliminary results with polychromatic light.

\section{PERSEE SHORT DESCRIPTION}
\label{sec:PERSEE SHORT DESCRIPTION}

This section provides the reader with a very brief description of PERSEE. Much more details can be obtained by the interested reader from ref{\cite{Cassaing08,Houairi08b,Jacquinod08,Henault10}}.

\subsection{PERSEE goals}
\label{sec:PERSEE goals}

The goal of PERSEE is not to reach the deepest possible nulling. Starting from a state of the art nuller of the 2006-2007 period, it is an experimental attempt to better master the system flowdown of the nulling requirements both at payload (instrument main optical bench) and platform levels (satellites). The balance of the constraints between those two levels is a key issue. The more disturbances the payload can face, the simpler the platforms are, the lower the cost. The general idea is hence to simplify as much as possible the global design and reduce the costs of a possible future space mission.

The detailed objectives have been described in ref{\cite{Cassaing08}}. Our main requirement is to reach a $10^{-4}$ nulling ratio in the [1.65-2.4]~\mum\;band (40\% spectral bandwidth) with a $10^{-5}$ stability over a 10~h time scale. Another important requirement is to be able to find and stabilize fringes which have an initial drift speed (as seen from the interferometer core) up to 150~\mum/s, as this can greatly simplify the relative metrology and control needs. We want to study and maximize the rejection of external disturbances introduced at relevant degrees of freedom of the optical setup by a disturbance module simulating various environments coming from the platform level.

\subsection{Optical setup general description}
\label{sec:setup description}

\begin{figure}
\begin{center}
\begin{tabular}{c}
\includegraphics[height=11.5cm, angle=270]{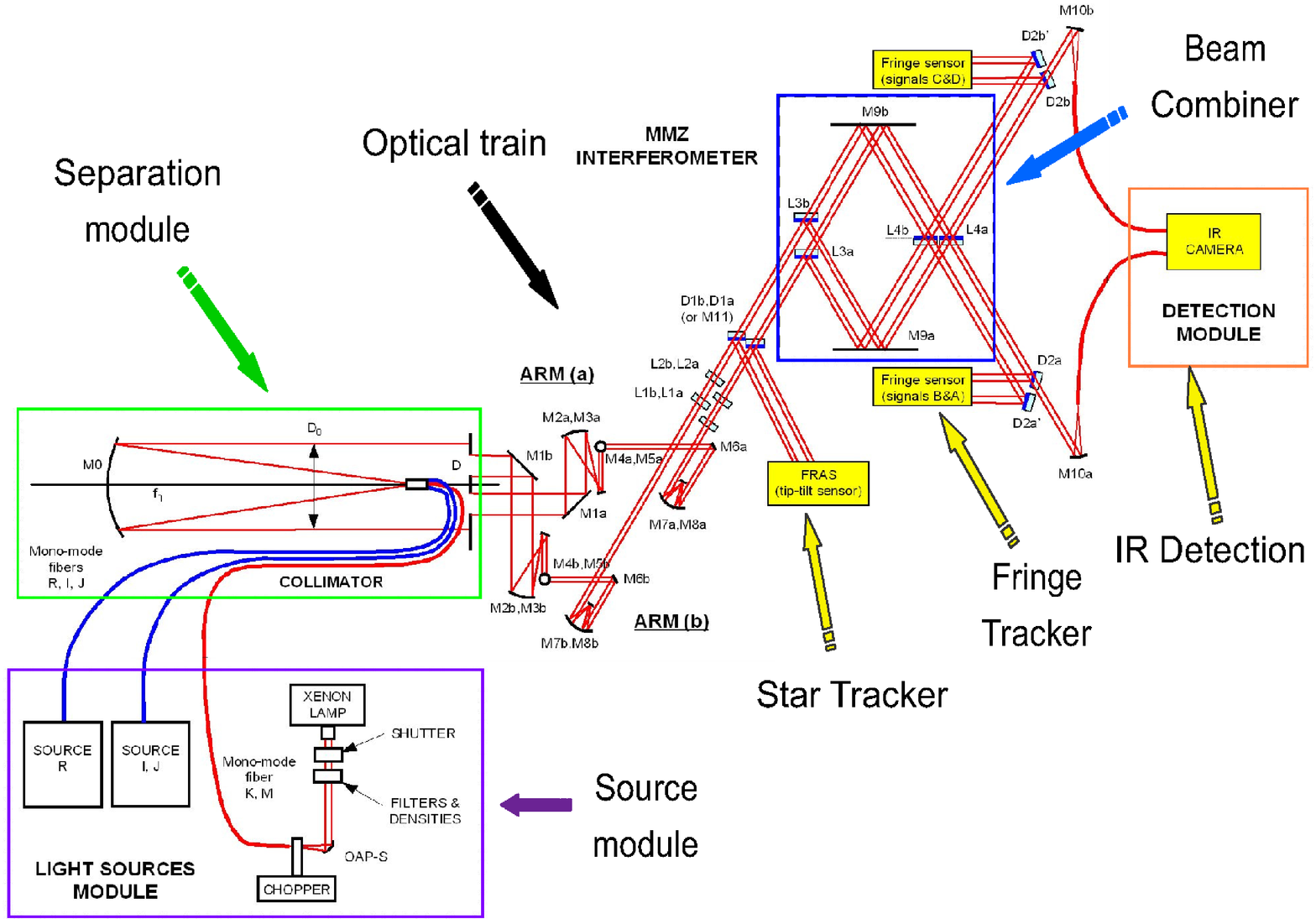}
\end{tabular}
\end{center}
\caption[setup]
{\label{fig:setup}
General setup of PERSEE in the final operational state.}
\end{figure}

Figure \ref{fig:setup} gives an overview of the optical layout. The source module combines various light sources in order to simulate a star light in a wide spectral range. All sources are injected into single-mode fibers which outputs are grouped together at the focus of a collimator. The I channel ([0.8-1]~\mum) is dedicated to the Fringe Sensor (FS) and the Field Relative Angle Sensor (FRAS). The J channel ([1-1.5]~\mum) is dedicated to the FS only. The nulling rate is measured in the K band ([1.65-2.4]~\mum). The 40~mm beams coming out of the separation module first encounter two 45\degre fold mirrors (M1) simulating siderostats. Accurate calibrated perturbations can be introduced at this level either in tip/tilt (few tens of mas resolution) or OPD (nm resolution). The optical train incorporates then afocal systems with $M=3$ magnification to reduce the beam size down to about 13~mm. M4 mirrors deviate the beam toward the -Z direction. Coupled with the M1 mirrors, they create a ``field reversal'' achromatic phase shift between the two arms{\cite{Serabyn01}}. Following, each arm has a cat's eye delay line (50~mm stroke and 10~nm resolution) and a 30\degre active M6 mirror acting both in tip/tilt and fine OPD (0.2~nm resolution). These active systems can all together generate various corrections covering the necessary range and resolution both in tip/tilt and OPD. They are used in dedicated control loops which use the FRAS camera for the tip/tilt and the FS for the OPD. L1 and L2, located just before the combining stage, are an optional phase shift compensator.

The combining stage is a kind of Modified Mach-Zehnder (MMZ)\cite{Serabyn01} with two inputs (a and b arms) and four outputs (I to IV). Due to some symmetry and optical tight tolerances, the setup guarantees that two outputs are achromatic. The four beam splitters have a trapezoidal geometry\cite{Jacquinod08} to avoid stray light. They have a specifically designed three-layer coating of about 200~nm thickness which allows to achieve appropriate reflection and transmission coefficients for both polarizations in the whole K band. The beam splitters are divided into two groups as compared to the classical setup{\cite{Serabyn01}}. This allow to translate L3a component without impacting the nulling output (output~III). The translation is adjusted so that the four outputs generate four points in the I and J fringes roughly with a  $\lambda/4$ spacing, which allows a spatial ABCD fringe tracking algorithm. The tip/tilt degrees of freedom of L3a are used to maximize the contrast in the FS A and C outputs (I and IV). This compact design allows to minimize differential optical path between FS and nulled output down to about 10~cm. This point is critical because it relaxes the stabilization requirements at system level as compared to other designs where differential paths can reach up to 1~m.

After the MMZ, dichroic plates separate the various spectral channels and direct them into appropriate detection chains via injection into optical fibers. The FS detection is made of a set of eight monopixels and multimode fibers. For the K channel, light is injected in a fluoride glass single mode fiber. The detector is a cooled infrared camera based on a Teledyne picnic matrix behind a dispersive element which allows to have polychromatic measurements of both dark and bright fringes at the same time. A calibrated optical attenuator is used in the bright output to optimize the dynamic of both outputs with respect to the detector well depth.

\section{RESULTS IN MONOCHROMATIC CONFIGURATION}
\label{sec:MONOCHROMATIC}

\subsection{Autocollimation setup}
\label{sec:autocol_setup}

\begin{figure}
\begin{center}
\begin{tabular}{c}
\includegraphics[height=7cm]{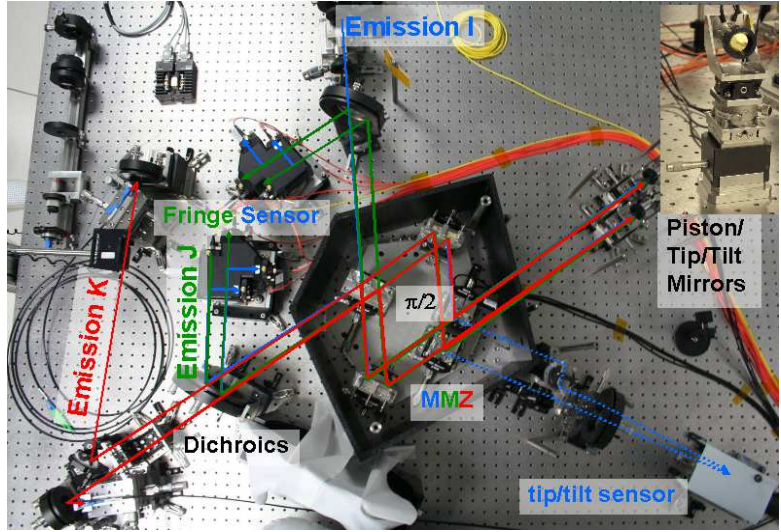}
\end{tabular}
\end{center}
\caption[schema_autocol]
{\label{fig:schema_autocol}
Autocollimation setup.}
\end{figure}

As PERSEE is a quite complex breadboard with many components, the integration was made progressively using many intermediate steps. The most important one at the start of the process was a so-called ``auto-collimation setup'' illustrated by figure \ref{fig:schema_autocol}. In this phase, we focused on the MMZ and the FS performances associated to a monochromatic nulled output measured by a single-pixel infrared detector. The several light sources were injected from some outputs and reflected back to inputs of the combination stage by the two M6 mirrors.

The I channel used a fibered DFB laser diode at 830~nm leading to a very long coherence length. It was injected in output II with a dedicated injection. The J channel used a SLED at 1320~nm with 40~nm spectral width leading to a coherence envelope of about 40~\mum.  It was injected in the reception fiber of output III. In the K channel, a DFB laser diode at 2320~nm was injected through a fluoride glass single mode fiber in output III. This setup had some drawbacks: by instance, the I flux in output III an IV were not balanced, contrary to the final setup. Furthermore, as an output was occupied by J source, the algorithm in this band was adapted to an ABC modulation.

\subsection{Calibration and thermal drift}
\label{sec:calibration}

\begin{figure}
\begin{center}
\begin{tabular}{c}
\includegraphics[height=5cm]{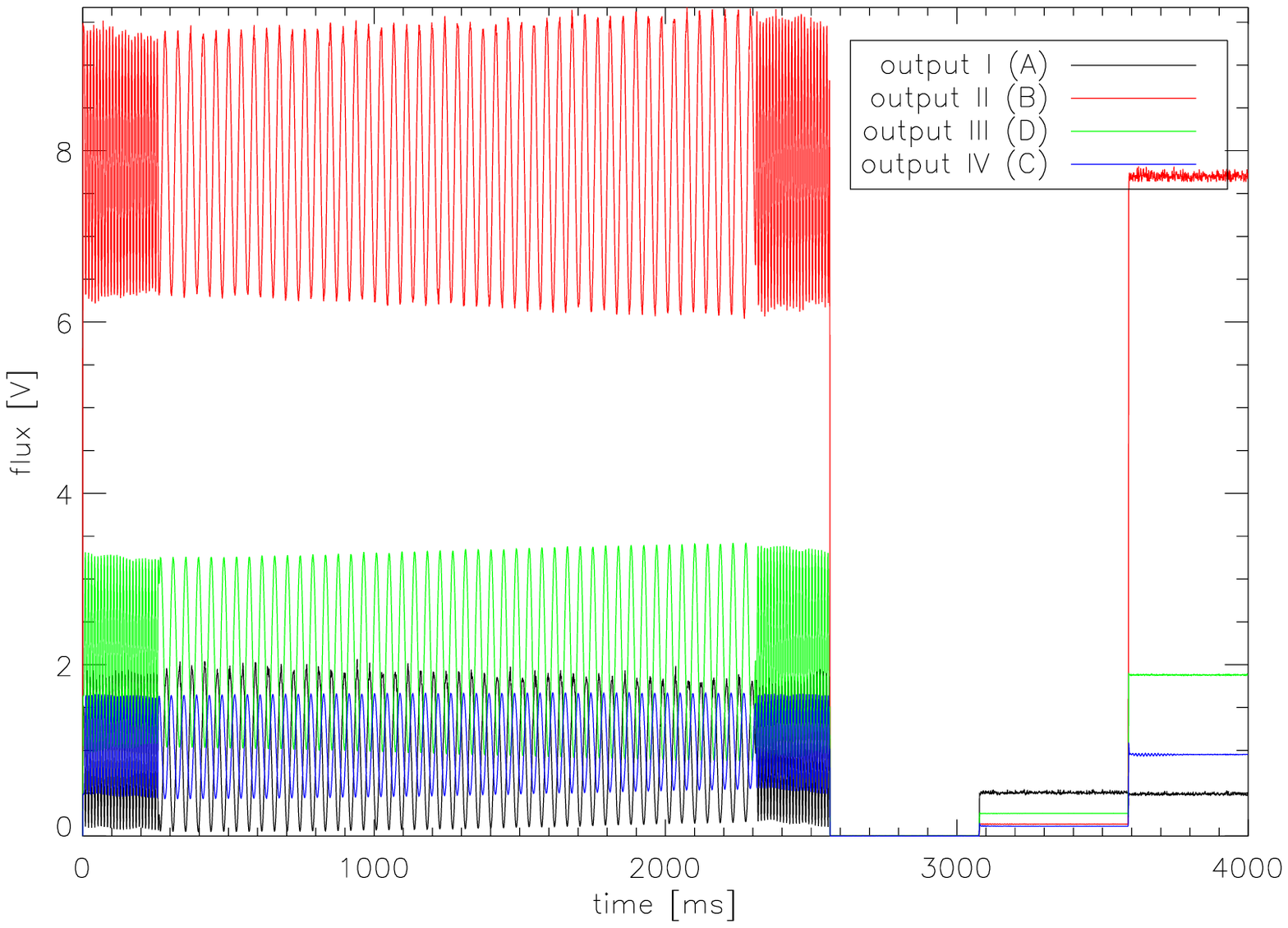}
\includegraphics[height=5cm]{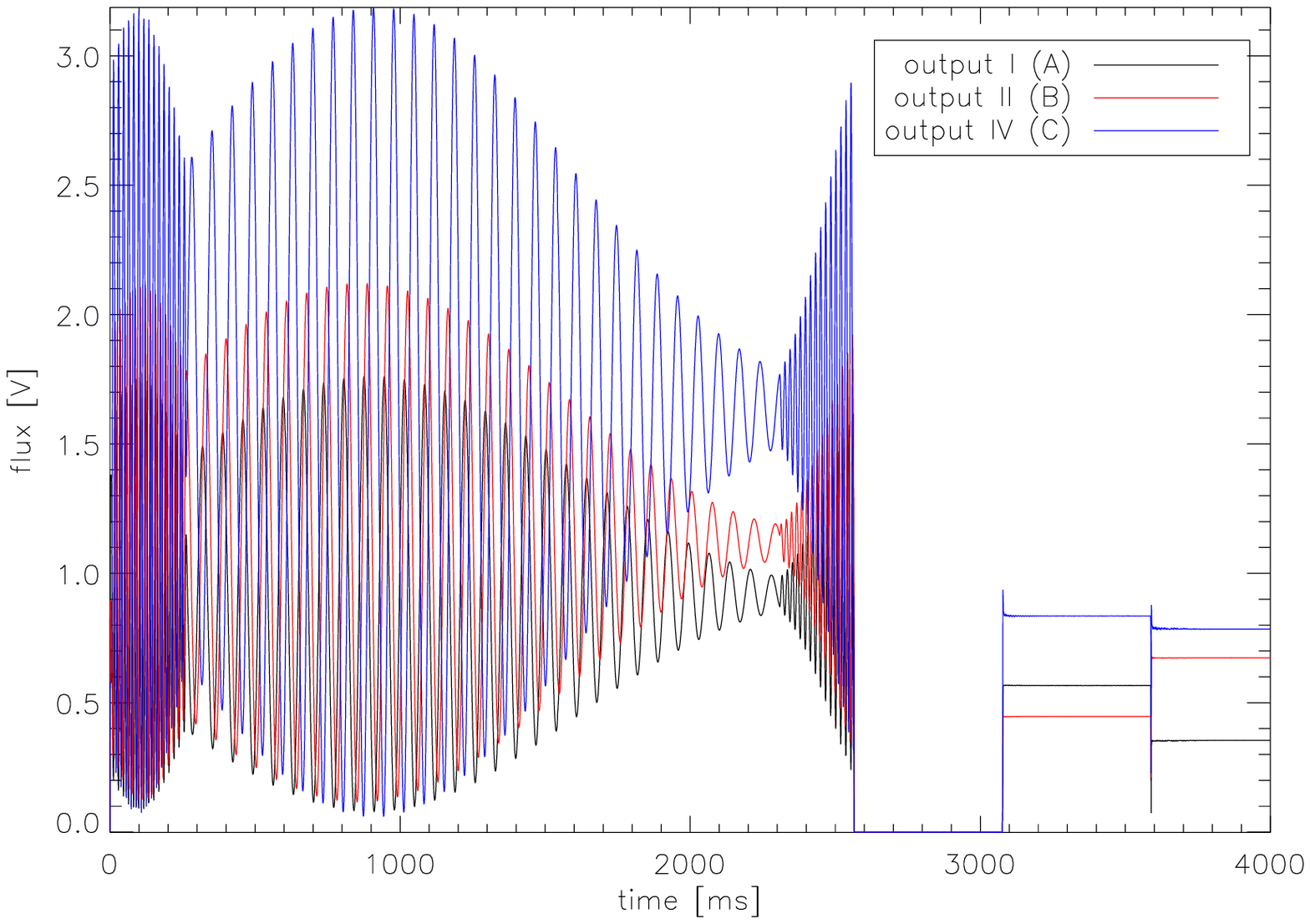}
\end{tabular}
\end{center}
\caption[calibration]
{\label{fig:calibration}
Signals obtained during calibration procedure, with 830~nm source (left) and 1320~nm source (right).}
\end{figure}

The main issue of spatial ABCD modulation is calibration. The system needs to be calibrated to perform a reliable measurement of OPD. Figure \ref{fig:calibration} presents a typical calibration file. 

\begin{figure}
\begin{center}
\begin{tabular}{c}
\includegraphics[height=5.3cm]{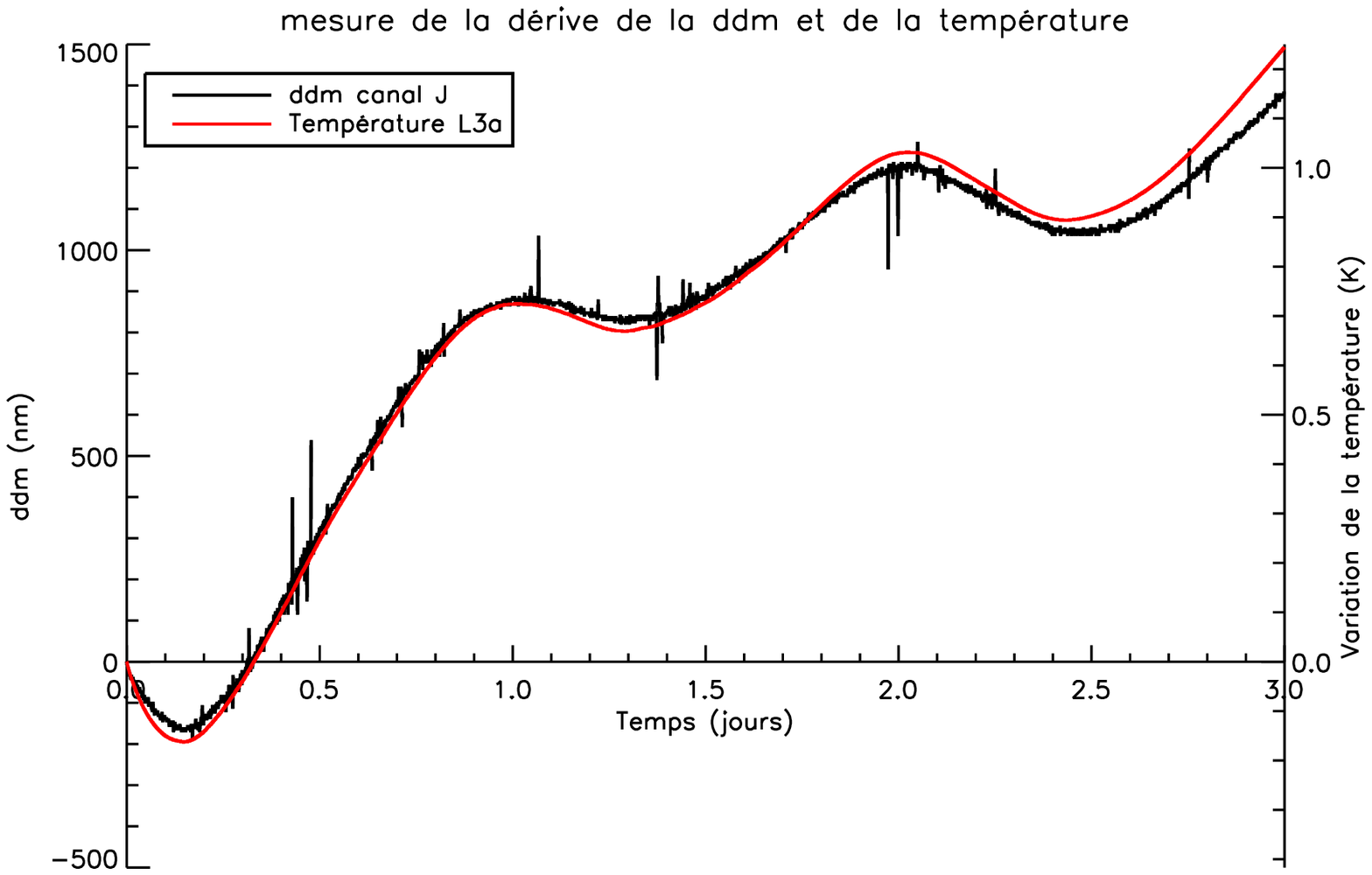}
\includegraphics[height=5.3cm]{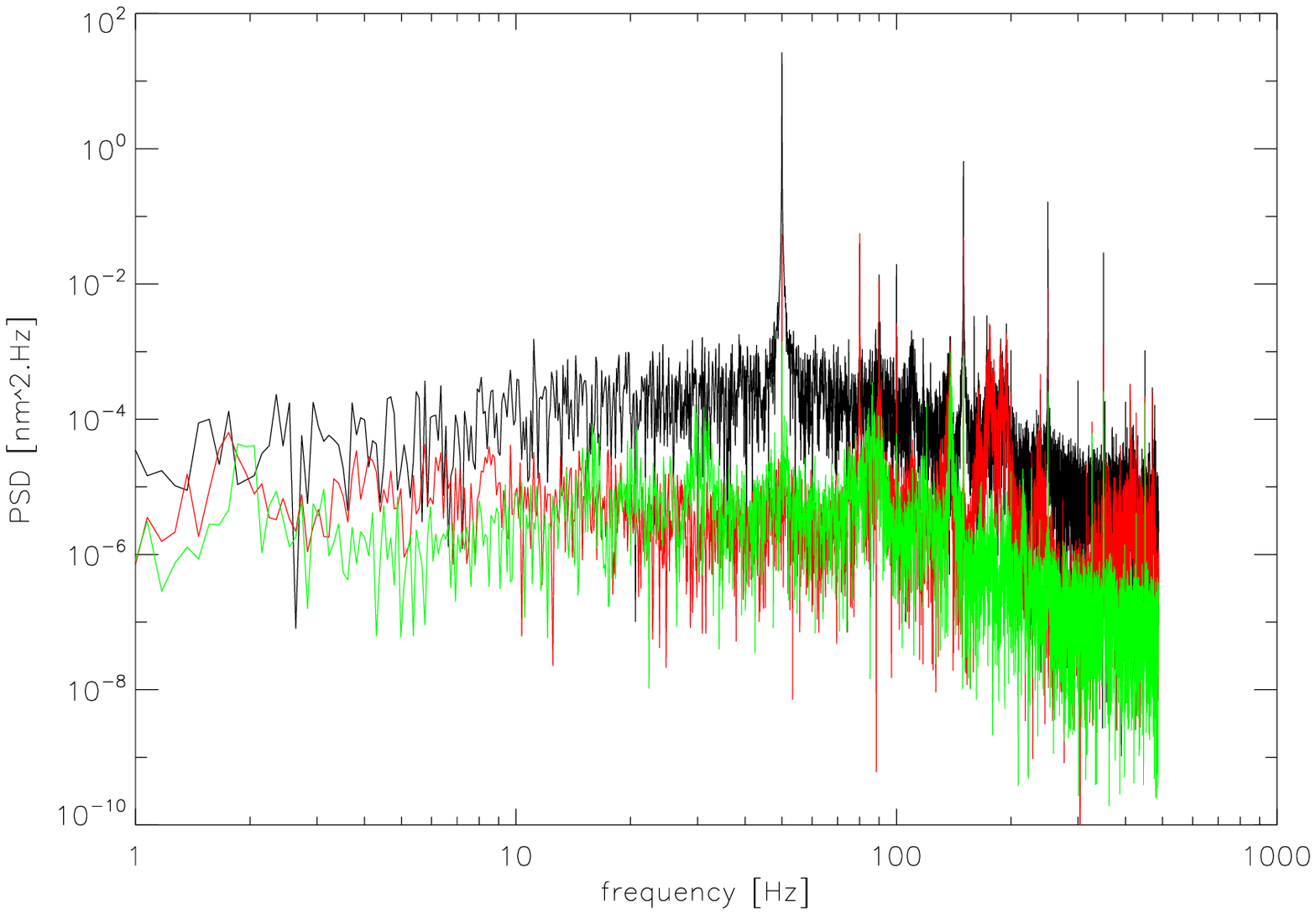}
\end{tabular}
\end{center}
\caption[opd_temp]
{\label{fig:opd_temp}
OPD between outputs IV and III and temperature variations versus time in days (left), and PSD of OPD measurements (left) w/ piezo strain gauges (black), w/o gauges (red), w/o gauges and other acoustic disturbances (green).}
\end{figure}

This file is obtained by applying an OPD ramp with the piezoelectric mirrors, generating interference fringes. This ramp is surrounded by fast OPD ramp to avoid voltage steps. Then the piezoelectric mirrors are commanded to an extreme tip/tilt position to perform a dark acquisition. Finally, we perform alternatively the same command to each mirror, to obtain the flux in each arms. With those data, it is possible to invert the modulation matrix. Using this matrix and the four outputs measurements, we can estimate in real-time the flux in each arm, the visibility and the OPD.In a stable environment, we do not need to calibrate frequently. But in reality, the coefficients of the demodulation matrix are sensitive to temperature variations. They are the main cause of recalibration. Figure \ref{fig:opd_temp} (left) compares the drift of outputs I and IV versus output II and III, and the temperature variations of a beam splitter inside the MMZ during 3 days. The correlation factor in that configuration is 1200~nm/K. In autocollimation, the light passes twice through the interferometer. So in the complete setup, that factor will be halved: 600~nm/K.

\subsection{Analysis of laboratory disturbances on OPD measurement}
\label{sec:analysis}

One of PERSEE goals is to cophase the two beams to a nanometric level, so it is very sensitive to laboratory disturbances, like acoustic or mechanical vibrations. Figure \ref{fig:opd_temp} (right) shows typically steps of noise reduction on the PSD of the OPD. The impact of piezo strain gauges is visible in a strong vibration at 50~Hz and a stronger white noise. By turning them off, we reduce significantly the noise, but there are still strong contributions around 80, 130 and 170~Hz. Those are mechanical vibrations: the M6 mirror assembly for the 80~Hz, and the MMZ for the 130 and 170~Hz. By deporting all electronic devices out of the room, and by turning off the A/C, we suppress every acoustic noise perturbator. Thereby, we reduce those mechanical vibrations. Nevertheless there are still some residues, that we will have to manage with the appropriate control law. The monochromatic results show that the two main contributors for the nulling rate are the photometric unbalance and the OPD noise.

\subsection{Best results in the autocollimation setup}
\label{sec:best_results}

In the best conditions, we cophased the two beams with a standard deviation of $\sigma_{OPD}=0.22$~nm in OPD (at 1~kHz during 10~s) and $\sigma_{tip/tilt}=60$~mas in tip/tilt (at 200~Hz during 10~s). In that setup, we obtained regularly that value of $\sigma$ in tip/tilt, and better than 0.4~nm in OPD. In tip/tilt, the performances are limited by measurement noise, contrary to the OPD performances, which are limited by laboratory disturbances. Data were acquired during 10~s, but results are very similar for 100~s time scales. 100~s is the typical duration between adjustments of satellite trajectories in the PEGASE mission framework.

For the monochromatic nulling rate at 2.32~\mum, we achieved an extinction of $N=3\cdot10^{-5}~\pm~3\cdot10^{-6}$ at 1~kHz during 100~s. That result is quite reproducible and is very comforting for the following polychromatic step.

\section{KALMAN FILTERING DESIGN AND TESTS}
\label{sec:KALMAN}

Vibrations are the main issue for nanometric cophasing systems. On PERSEE, we want to minimize the impact of laboratory environment, in order to accurately analyze effects of calibrated disturbances that we will inject later. Passive filtering methods exist to reduce vibrations, but they fail to filter them entirely. So we considered an active method, already studied\cite{Petit08} for the XAO instrument SPHERE at ONERA. This method was adapted to our problem, and is currently tested on the bench. Linear Quadratic Gaussian (LQG) control laws can provide an optimal correction of vibrations, in the sense of residual phase minimum variance. Nevertheless, they require an identification procedure, which provides regularly the updated characteristics of the disturbances.

\subsection{LQG vibration control}
\label{sec:LQG}

Unlike AO, we will consider only the first three Zernike modes: piston, tip and tilt. However, the model remains the same. We consider that active OPD/tip/tilt mirrors provide a linear and instantaneous response, constant over a frame period T. So the correction phase $\phi^{cor}_{n-1}$ during the time period $[(n-2)T,(n-1)T]$ is:
\begin{equation}\label{eq:phi-corr}
\phi^{cor}_{n-1}=Nu_{n-2},
\end{equation}
where $N$ is the influence matrix and $u$ the command sent to the actuators. We consider a two frame period delay system. We assume that the disturbance signal is the sum of uncorrelated disturbances, like atmospheric turbulence and vibrations. Thus, during the time interval $[(n-1)T,nT]$, the disturbed phase $\phi^{dis}_n$ is:
\begin{equation}\label{eq:phi-dis}
\phi^{dis}_n=\phi^{turb}_n+\sum_{N_vib}\phi^{vib,i}_n.
\end{equation}

The FS and the FRAS provide a signal $y_n$ integrated over the time interval $[(n-2)T,(n-1)T]$. This signal is used to compute the command $u_n$ applied during the time interval $[nT,(n+1)T]$. $y_n$ is defined by:
\begin{equation}\label{eq:y-n}
y_n=D(\phi^{dis}_{n-1}-\phi^{cor}_{n-1})+w_n.
\end{equation}

In AO, the matrix $D$ characterizes the wavefront sensor, so here it characterizes the FS and the FRAS. $w_n$ is the zero-mean white Gaussian measurement noise. Its covariance matrix is noted $\Sigma_w$.
We describe each disturbance as a second-order auto-regressive model (AR2):
\begin{alignat}2\label{eq:phi-tur}
\phi^{tur}_{n+1}&=A_1^{tur}\phi^{tur}_{n}+A_2^{tur}\phi^{tur}_{n-1}+v^{tur}_{n},\\
\phi^{vib,i}_{n+1}&=A_1^{vib,i}\phi^{vib,i}_{n}+A_2^{vib,i}\phi^{vib,i}_{n-1}+v^{vib,i}_{n}.
\end{alignat}

$A_1$ and $A_2$ are matrices defining the dynamic behavior of the disturbance. $v_n$ is a zero-mean white Gaussian noise with a covariance matrix $\Sigma_v$.

We can recast the equations \ref{eq:phi-tur} in a matrix function:
\begin{equation}\label{eq:phi-mat}
\phi^{glob}_{n+1}=\begin{pmatrix}
\phi^{tur}_{n+1}\\
\phi^{tur}_{n}\\
\phi^{vib,1}_{n+1}\\
\phi^{vib,1}_{n}\\
\vdots
\end{pmatrix}=\begin{pmatrix}
A^{tur}_1 & A^{tur}_2 & 0 & 0 & \cdots\\
Id & 0 & 0 & 0 & \cdots\\
0 & 0 & A^{vib,1}_1 & A^{vib,1}_2 & \cdots\\
0 & 0 & Id & 0 & \cdots\\
\vdots  & \vdots  & \vdots  & \vdots  & \ddots
\end{pmatrix}\begin{pmatrix}
\phi^{tur}_{n}\\
\phi^{tur}_{n-1}\\
\phi^{vib,1}_{n}\\
\phi^{vib,1}_{n-1}\\
\vdots
\end{pmatrix}+\begin{pmatrix}
v^{tur}_{n}\\
0\\
v^{vib,1}_{n}\\
0\\
\vdots
\end{pmatrix}=A^{glob}\phi^{glob}_{n}+v^{glob}_{n}.
\end{equation}

If $x_n$ is the state vector of the system:
\begin{equation}\label{eq:x-n}
x_n=\begin{pmatrix}
\phi^{glob}_n\\
u_{n-1}\\
u_{n-2}
\end{pmatrix},
\end{equation}
the whole process corresponds to the following time-invariant state-space model, similar to a classical process control system:
\begin{alignat}2\label{eq:state-eq}
x_{n+1}&=\begin{pmatrix}
A^{glob} & 0 & 0\\
0 & 0 & 0\\
0 & Id & 0
\end{pmatrix}x_{n}+\begin{pmatrix}
0\\
Id\\
0
\end{pmatrix}u_{n}+\begin{pmatrix}
v^{glob}_n\\
0\\
0
\end{pmatrix}&&=Ax_n+Bu_n+\nu_n,\\
y_n&=\begin{pmatrix}
0 & D & 0 & D & \cdots & 0 & -DN
\end{pmatrix}x_n+w_n&&=Cx_n+w_n.
\end{alignat}

\subsection{Steps of the asymptotic gain Kalman filter}
\label{sec:steps}

An asymptotic gain Kalman filter can be described in 3 different steps :
\begin{alignat}2
\text{innovation computing:  }&&\tilde{y}_{n|n-1}&=y_n-\hat{y}_{n|n-1}=Cx_n+w_n-C\hat{x}_{n|n-1}=C\tilde{x}_n+w_n, \\
\text{update:  }&&\hat{x}_{n|n}&=\hat{x}_{n|n-1}+H_\infty \tilde{y}_{n|n-1}, \\
\text{prediction:  }&&\hat{x}_{n+1|n}&=A\hat{x}_{n|n}+Bu_n.
\end{alignat}

$H_\infty$, the Kalman gain, is calculated outside the loop, using the following equation:
\begin{equation}\label{eq:kalman_gain}
H_\infty=\Sigma_\infty C^t\left(C\Sigma_\infty C^t+\Sigma_w \right)^{-1},
\end{equation}
with $\Sigma_\infty$ the asymptotic solution of Riccati's algebraic equation:
\begin{equation}\label{eq:riccati}
\Sigma_\infty=A\Sigma_\infty A^t+\Gamma \Sigma_\nu \Gamma^t-A\Sigma_\infty C^t\left(C\Sigma_\infty C^t+\Sigma_w \right)^{-1}C\Sigma_\infty A^t.
\end{equation}

Those equations can be quite onerous in computation time, but they are done outside the time-critical loop, and do not need to be actualized often.
In the same time, the parameters of matrix $A$ are determined outside the loop, using pseudo open-loop data, obtained during closed loop with the following equation:
\begin{equation}
y_{pseudo\;open-loop,n}=y_n+D\! Nu_{n-2}
\end{equation}

\subsection{Results of simulations}
\label{sec:simulations}

\begin{figure}
\begin{center}
\begin{tabular}{c}
\includegraphics[height=5.0cm]{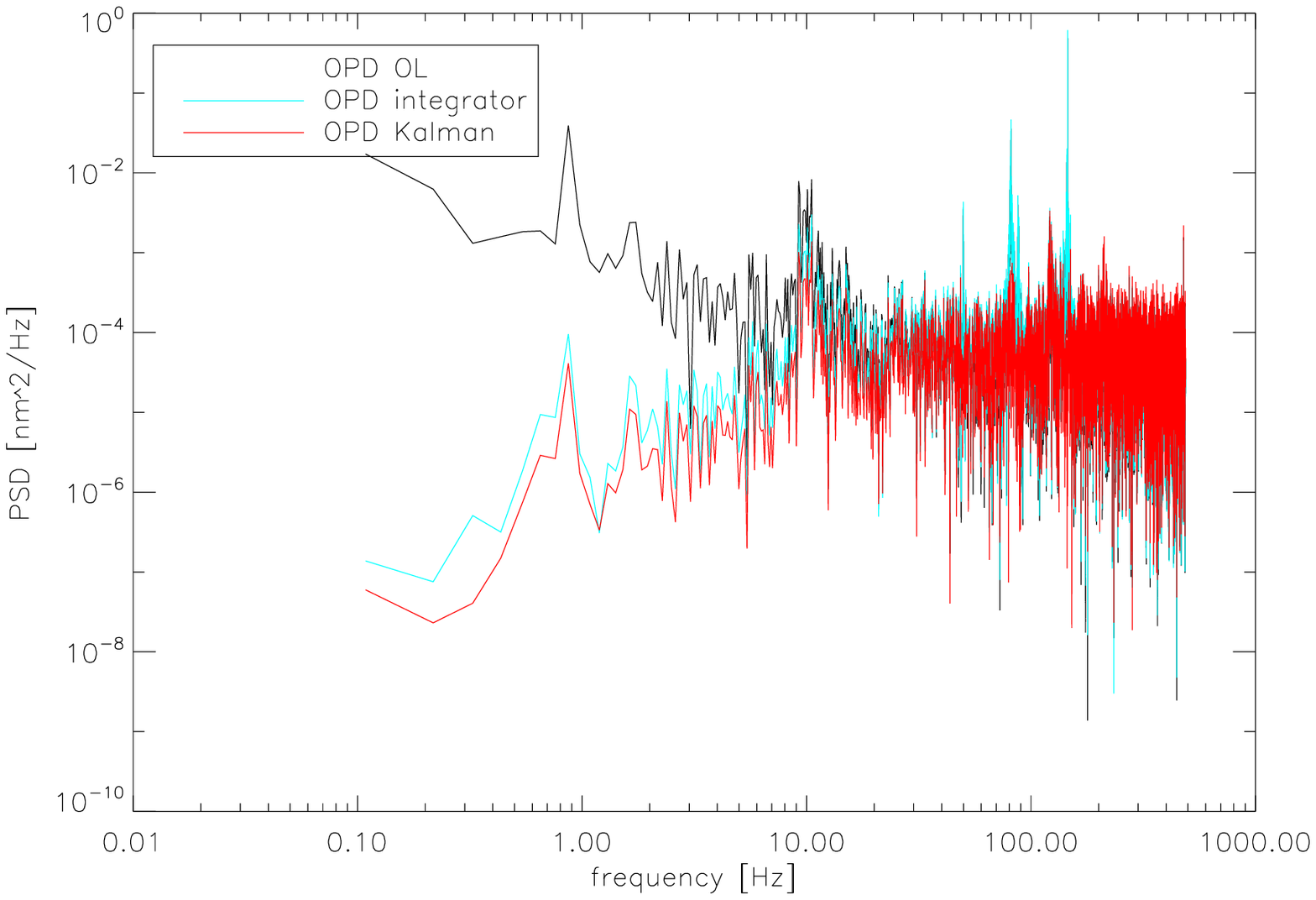}
\includegraphics[height=5.0cm]{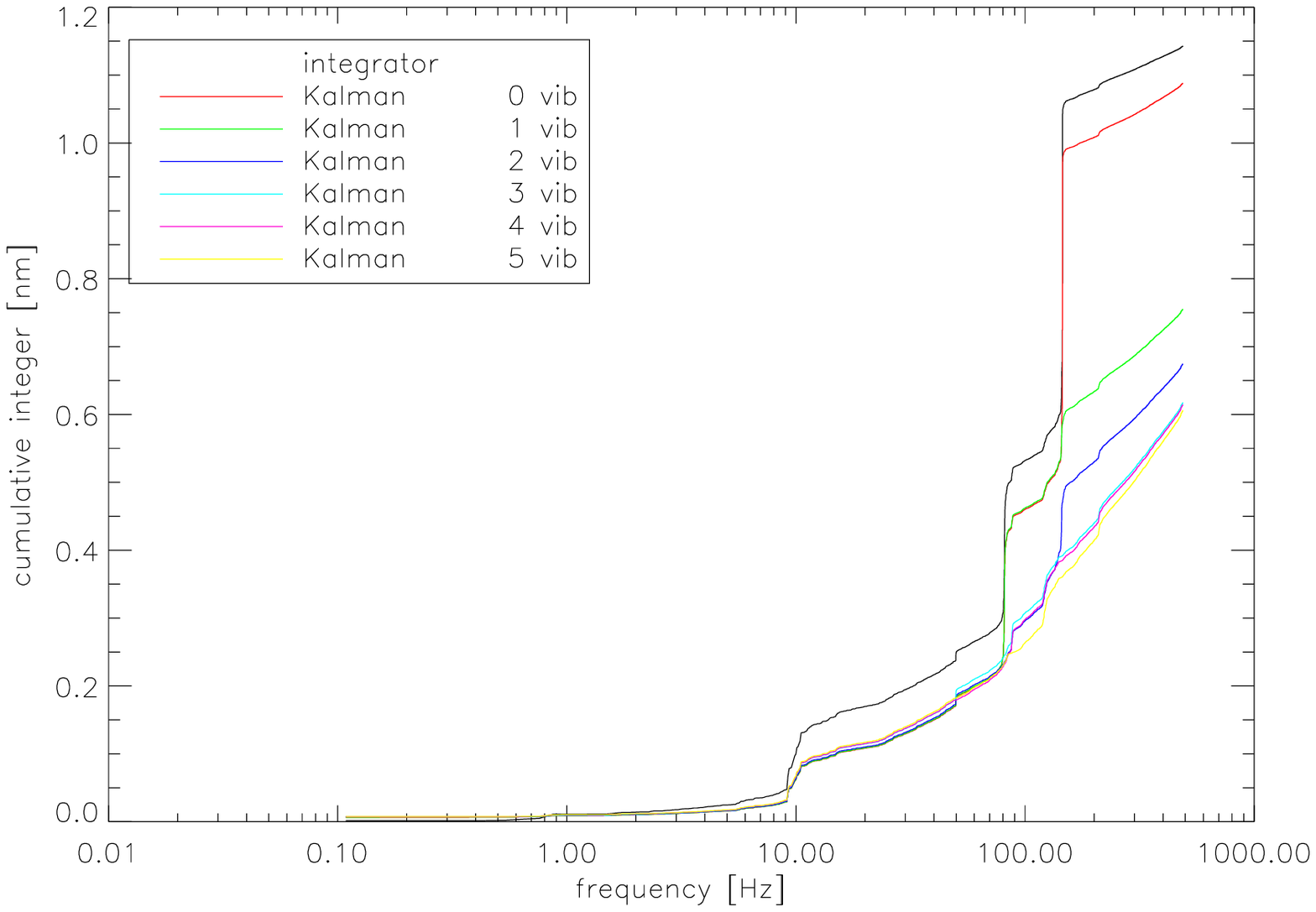}
\end{tabular}
\end{center}
\caption[kalman]
{\label{fig:kalman}
Closed loop simulations on real OPD data. Power spectral density with a classical integrator and with a Kalman filter (left), and cumulative integral of PSD for a classical integrator and Kalman filter, with increasing number of corrected vibrations.}
\end{figure}

Simulations of the LQG control were performed, with real open-loop measurement of OPD. Figure \ref{fig:kalman} shows simulation results, comparing a classical integrator to the Kalman filter. The simulator takes into account a two frame period delay. In that case, the integrator filter has an amplified zone around the cutoff frequency: 100~Hz. So the mechanical vibrations, which are between 80 and 170~Hz are amplified. Those vibrations can be very disturbing, because they can add several nanometers on the OPD noise. They are efficiently reduced by the Kalman filter. Even the turbulence contribution is better managed. This result is clearly visible on figure \ref{fig:kalman} (right). As described in \ref{sec:LQG}, the different vibration contributors are managed independently, so it is possible to analyze every contribution of identified vibrations. Even with turbulence identification only, the Kalman filter is more efficient than the integrator, but the most interesting effect is on identified vibrations. Only three corrected vibrations are enough to reduce the OPD residue of a factor of two.

\subsection{First application on PERSEE}
\label{sec:application}

\begin{figure}[t]
\begin{center}
\begin{tabular}{c}
\includegraphics[height=5.0cm]{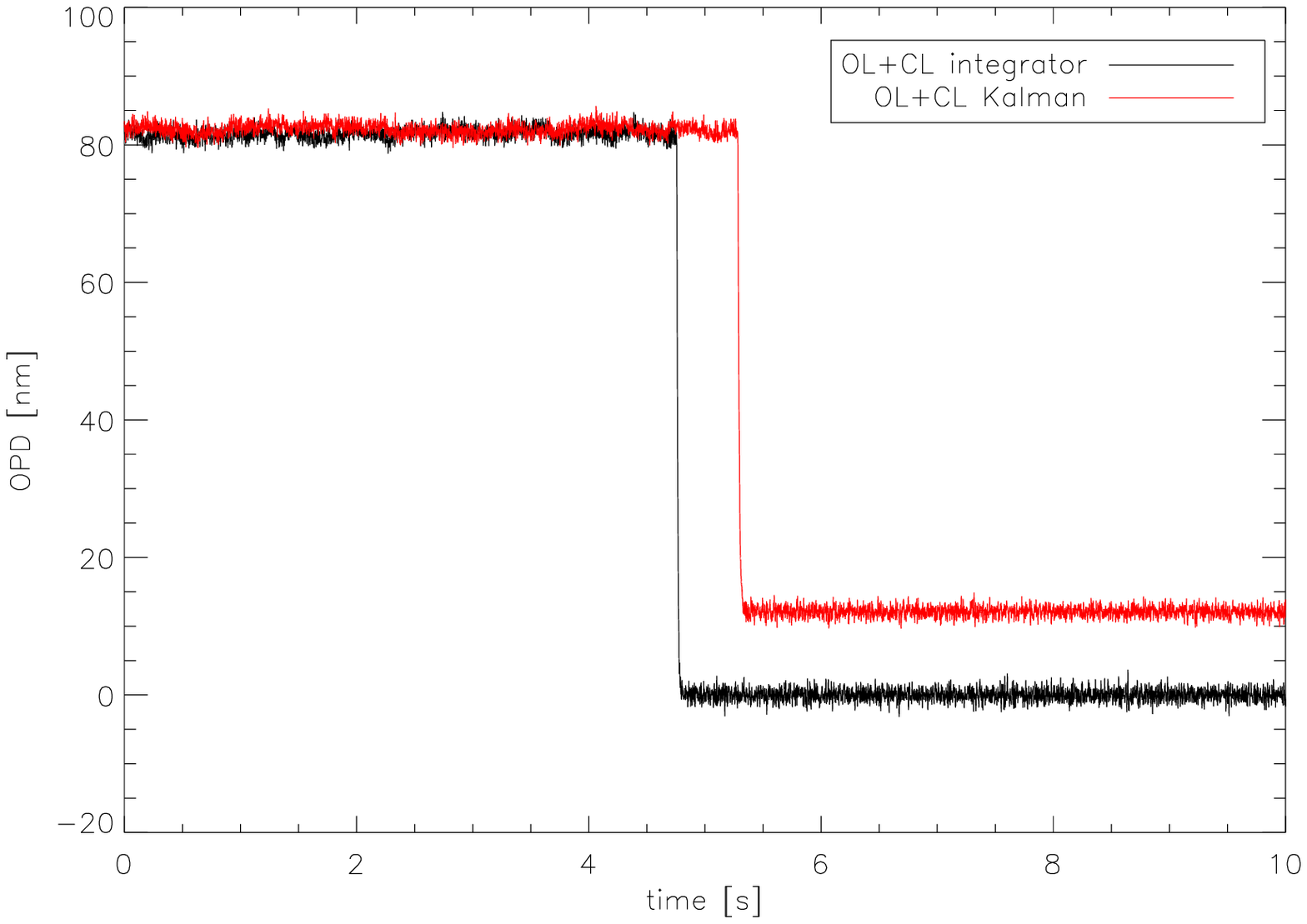}
\includegraphics[height=5.0cm]{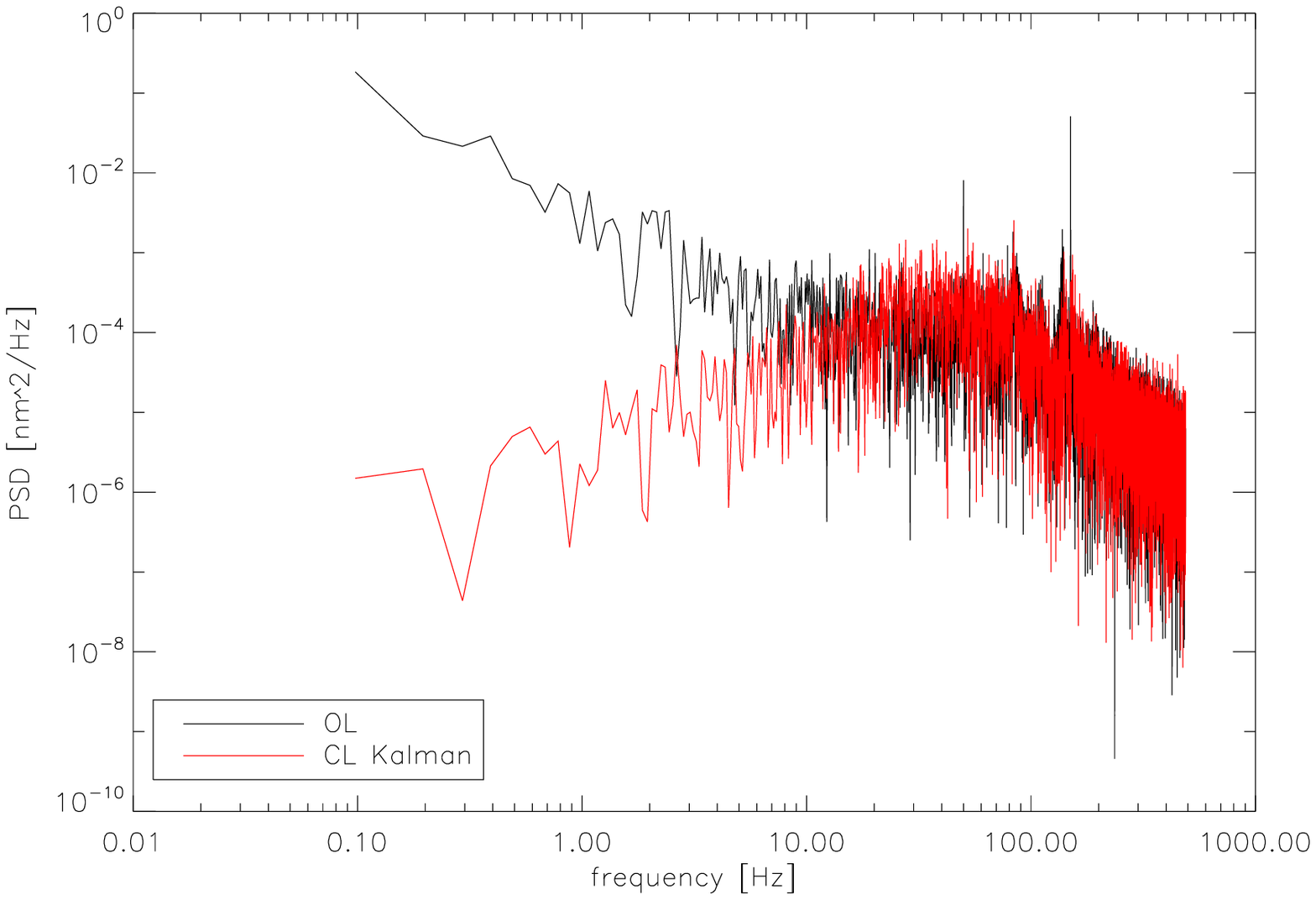}
\end{tabular}
\end{center}
\caption[int+kal]
{\label{fig:int+kal}
First tests of the Kalman filter on the bench. Temporal loop closure (left) with a classic integrator and a Kalman filter, and PSD comparison with open loop (OL) and closed loop (CL) with the two filters.}
\end{figure}

The Kalman filter is currently tested on the bench, to correct vibrations on OPD only. Disturbance identification is made with open-loop data, and we selected to correct 3 vibrations. Figure \ref{fig:int+kal} (left) shows results of the first loop closure. On the temporal data, we can see that when we close the loop with the Kalman filter, the OPD does not variate around zero, but around a non-zero value. That mean value is due to the fact that we assume a zero-mean turbulence, which is not the case. It can be easily removed in a second step. In Figure \ref{fig:int+kal} (right), we can see on the PSD that the vibration pics at 50, 130 and 150~Hz are clearly reduced. Other tests should confirm the efficiency of the Kalman filter.

\section{FIRST RESULTS IN POLYCHROMATIC CONFIGURATION}
\label{sec:POLYCHROMATIC}

\subsection{Setup}
\label{sec:setup}

\begin{figure}
\begin{center}
\begin{tabular}{c}
\includegraphics[height=3cm]{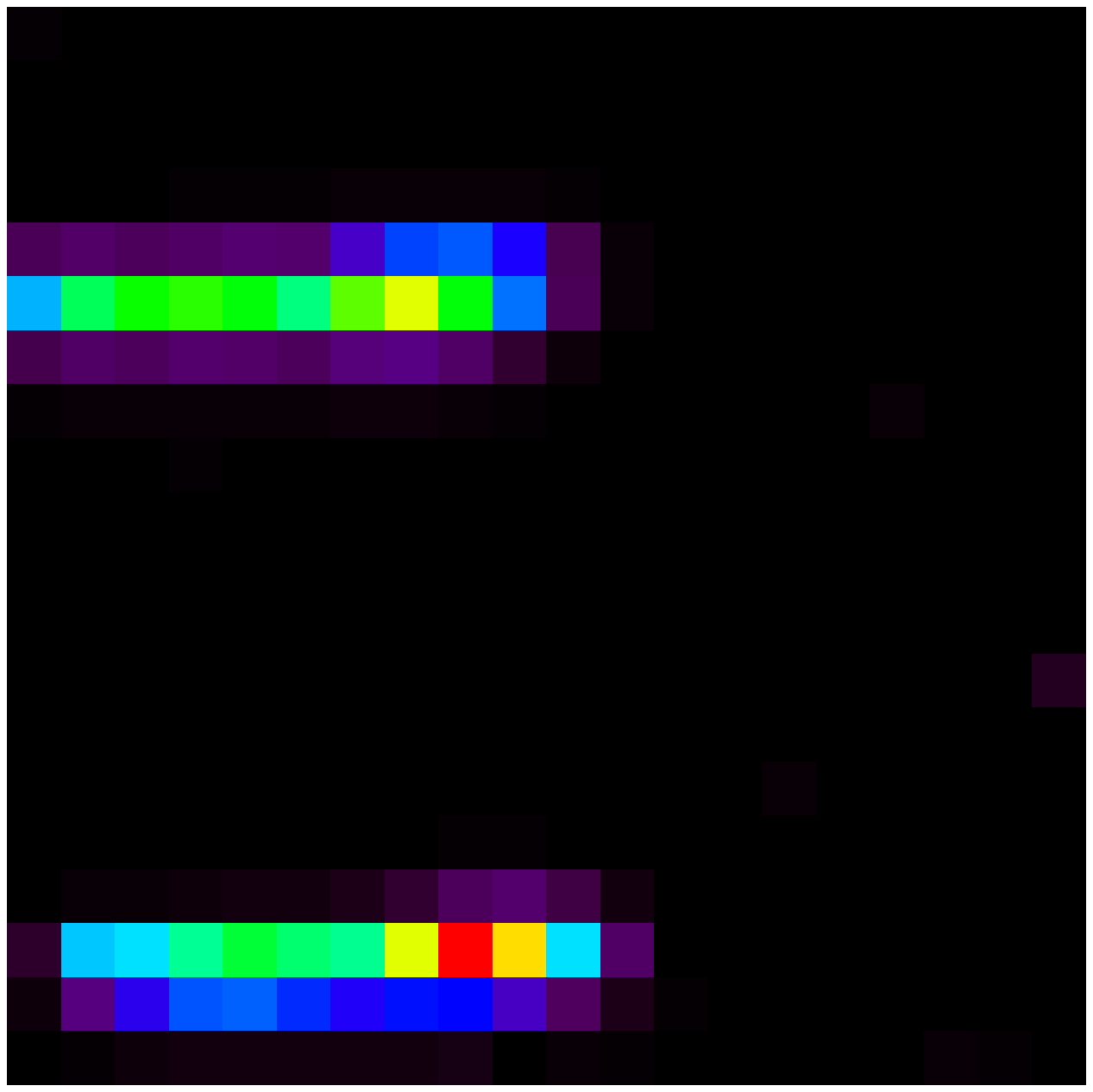}
\includegraphics[height=3cm]{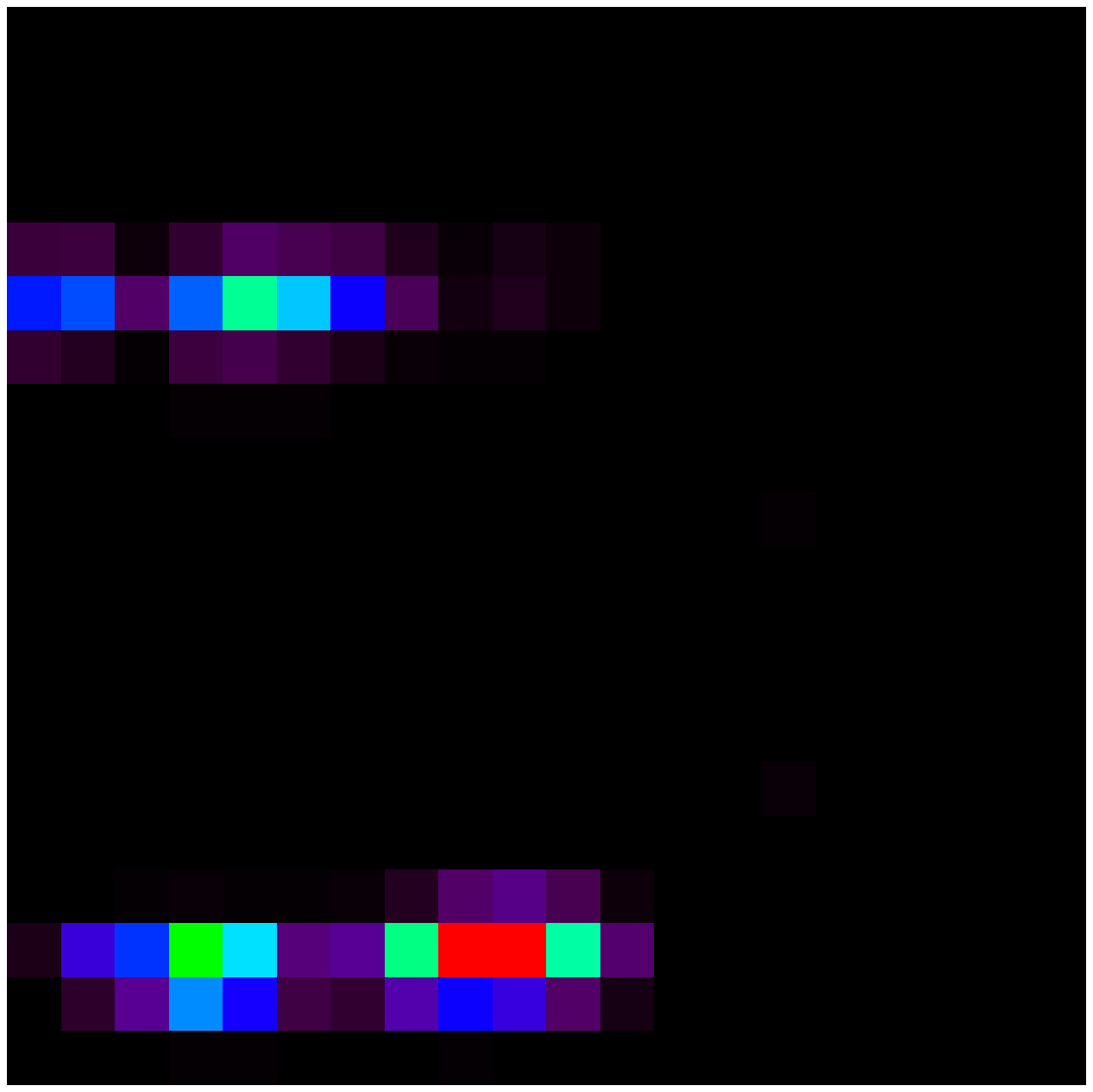}
\includegraphics[height=3cm]{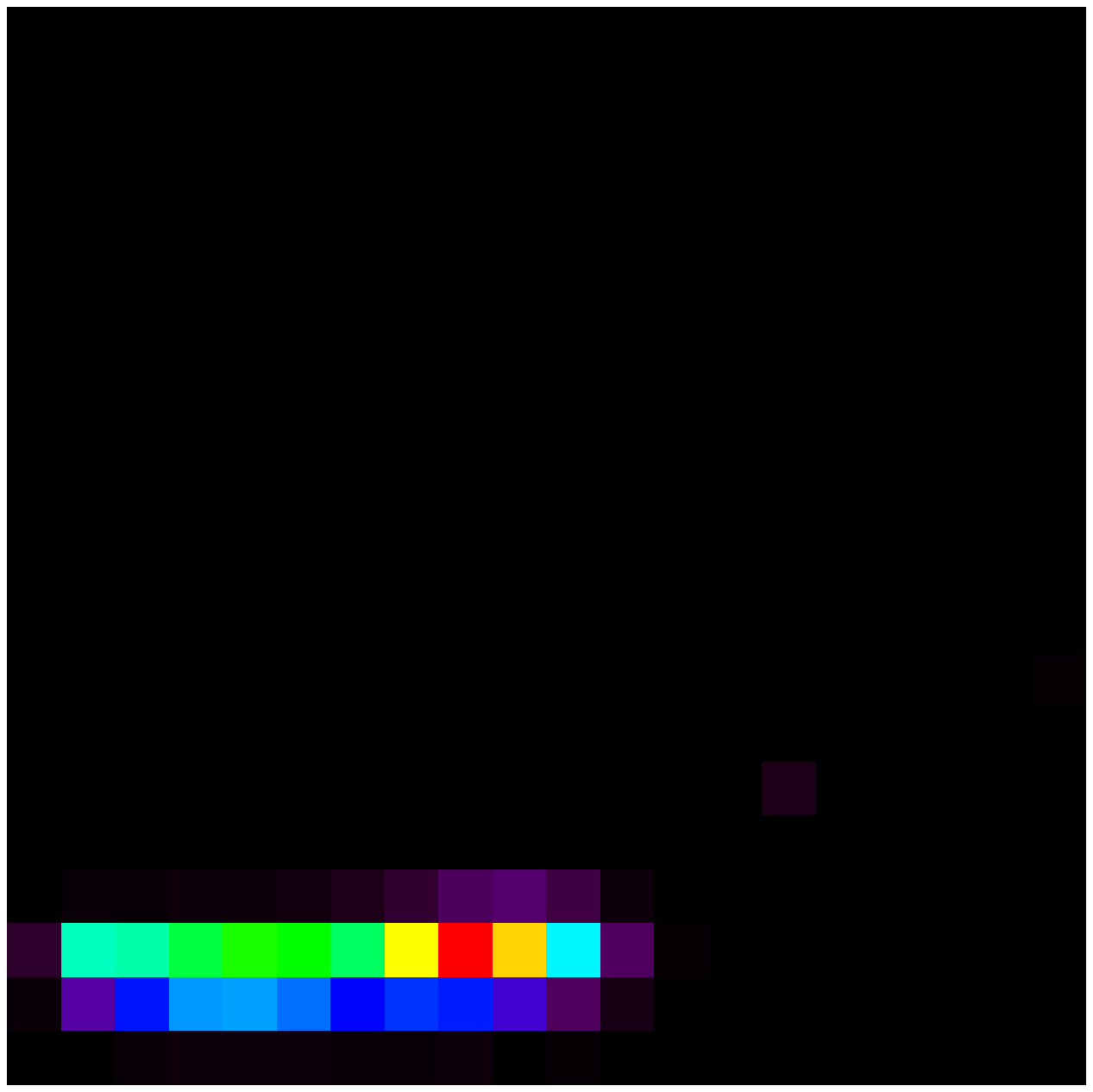}
\end{tabular}
\end{center}
\caption[im_cam]
{\label{fig:im_cam}
Images of the camera: one arm only (left), interferences at $OPD\approx10~\mum$ (center), interferences at $OPD=0$ (right). The upper spot (rows 13,14,15) is the output III (dark output) and the lower spot (rows 1,2,3) is the output II (bright output).}
\end{figure}

These tests were performed in a nearly complete configuration of the bench. Only afocal systems are missing. They were conducted to validate a supercontinuum source that will eventually replace the black body and the two lasers described in section \ref{sec:autocol_setup}. At that time, the optical bench was not optimized, so the results should be improved. The images shown in figure \ref{fig:im_cam} were obtained by the low noise nitrogen-cooled camera described in section \ref{sec:setup description}. The upper spot is output III, corresponding to the dark output, imaged through a bi-prism which disperses wavelengths. The lower spot is output II, corresponding to the bright output. Each spot is dispersed horizontally (1.6 to 2.3~\mum on 11 pixels).

\begin{figure}
\begin{center}
\begin{tabular}{c}
\includegraphics[height=5.0cm]{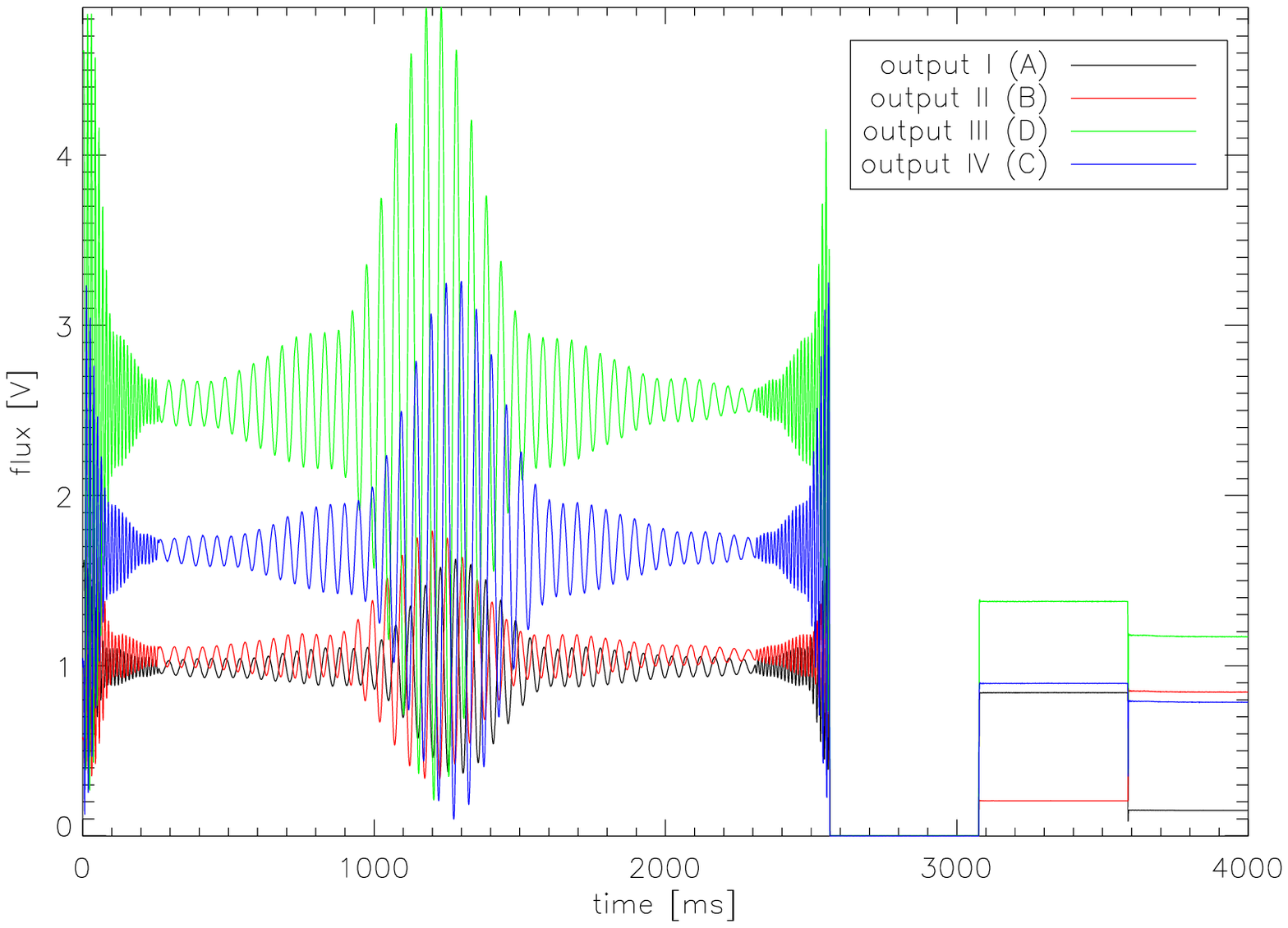}
\includegraphics[height=5.0cm]{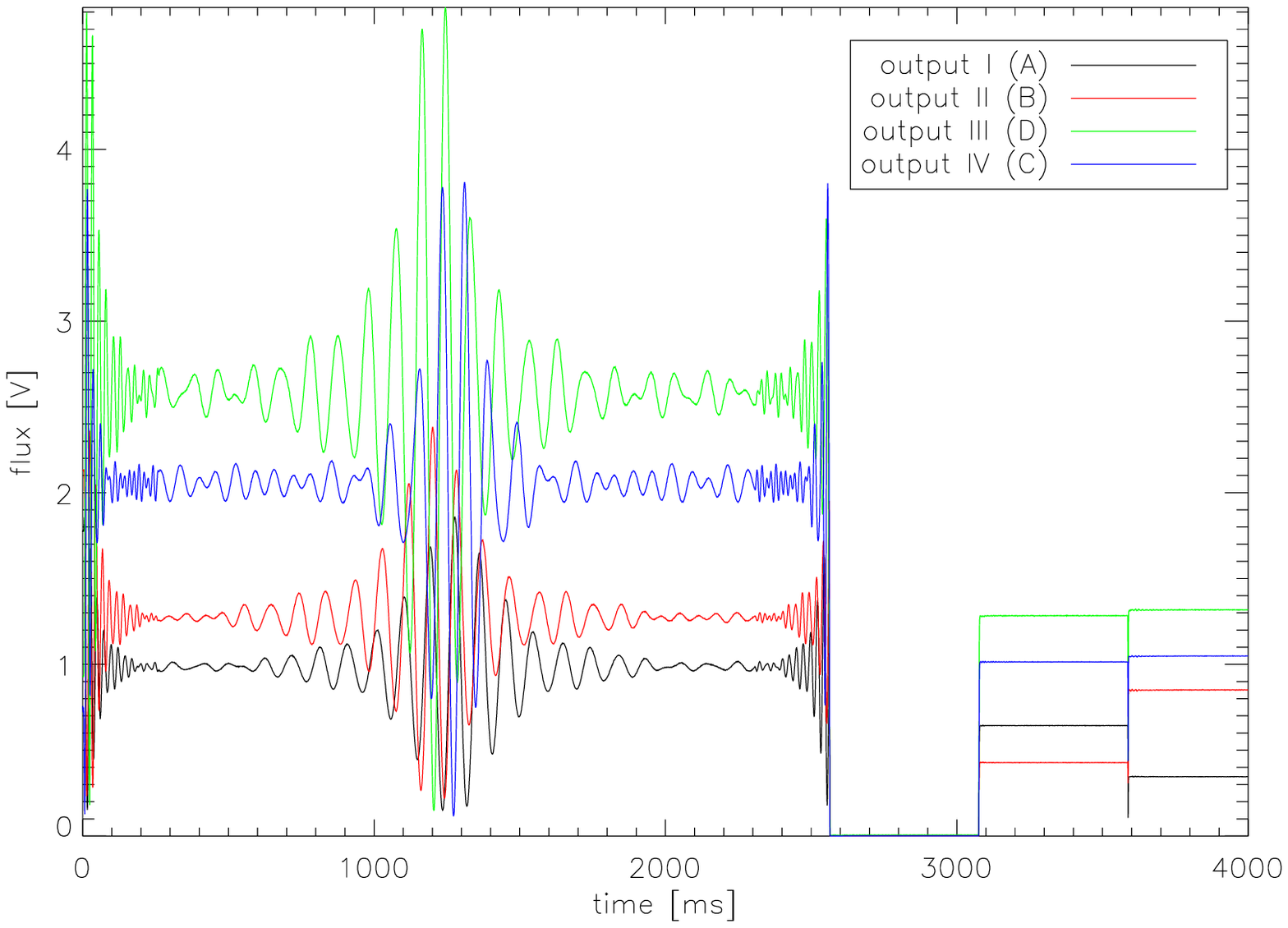}
\end{tabular}
\end{center}
\caption[calib_fianium]
{\label{fig:calib_fianium}
Calibration file with the supercontinuum, in I band (left) and J band (right).}
\end{figure}

In that setup, the supercontinuum source is used for both metrology and nulling measurement. The range of metrology is then quite limited by the coherence lengths of the two spectral bands, shorter than those of the SLED and the DFB diode previously used in section \ref{sec:MONOCHROMATIC}. Figure \ref{fig:calib_fianium} is a calibration file obtained with the supercontinuum source. Compared to figure \ref{fig:calibration}, we can see the shorter coherence length of the spectral bands.

\subsection{Photometry}
\label{sec:photometry}

\begin{figure}
\begin{center}
\begin{tabular}{c}
\includegraphics[height=6.0cm]{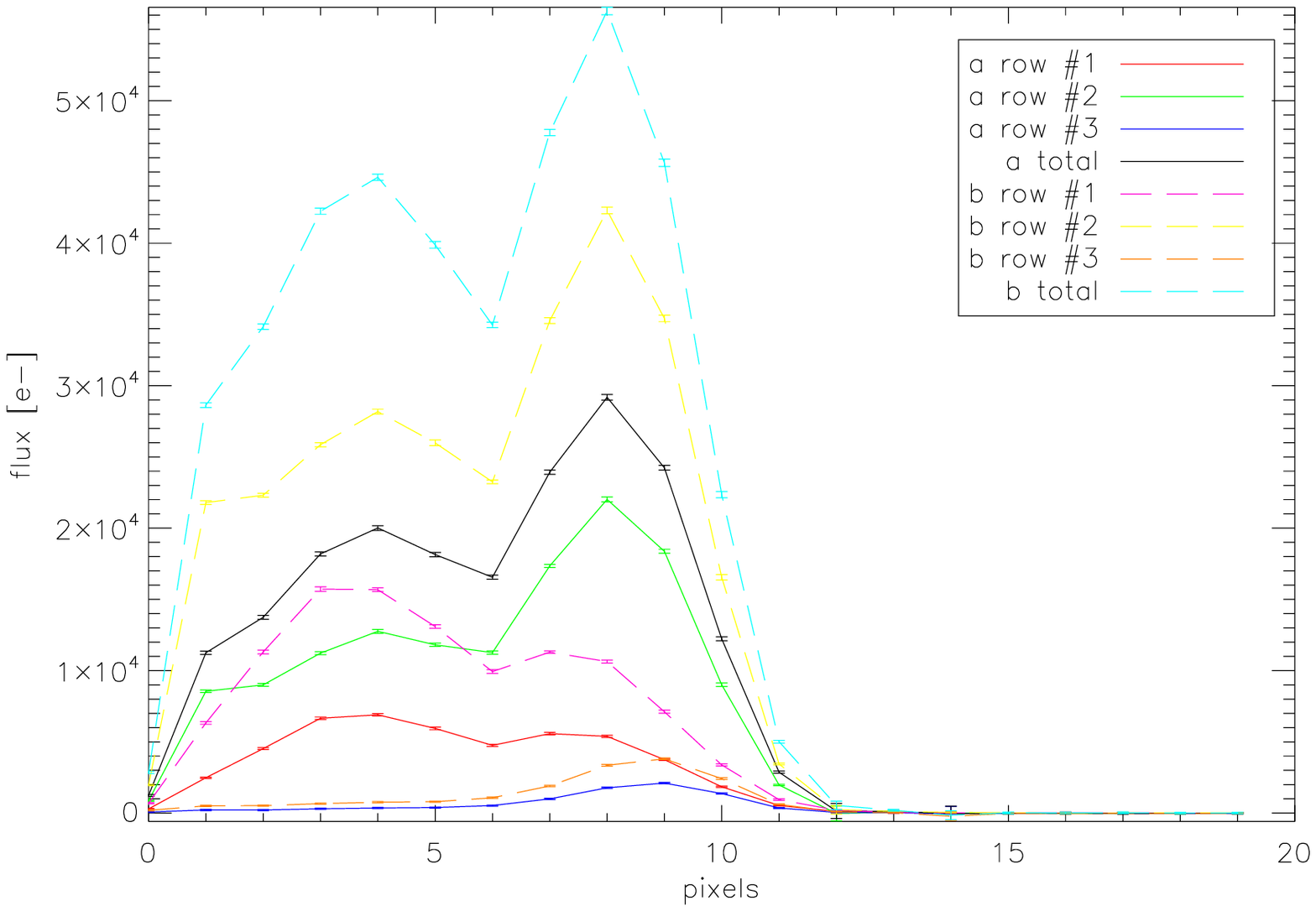}
\includegraphics[height=6.0cm]{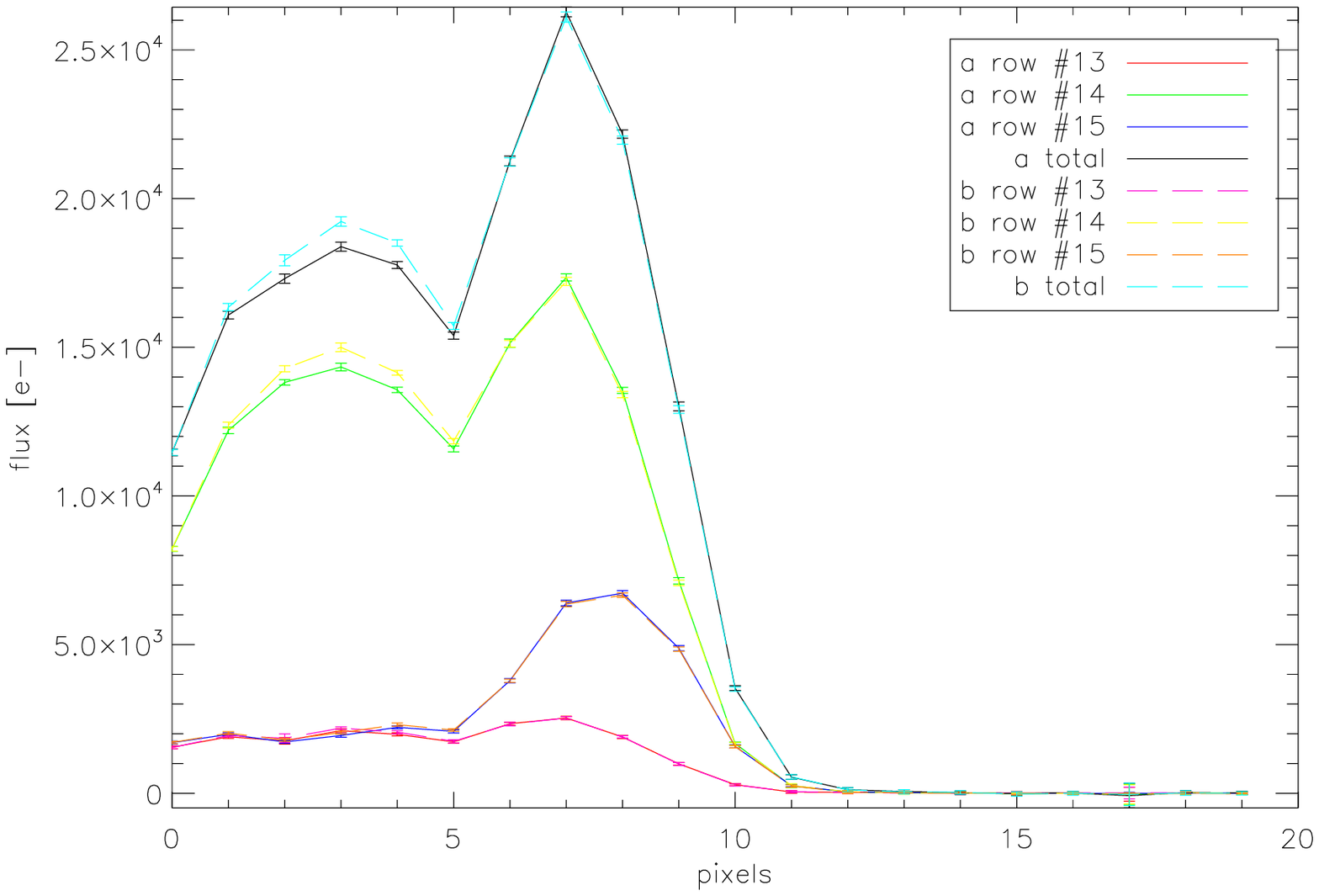}
\end{tabular}
\end{center}
\caption[spII-III]
{\label{fig:spII-III}
Flux in each arm (a and b) in output II (left) and III (right).}
\end{figure}

First, we acquire the flux in each arm. Figure \ref{fig:spII-III} shows those flux, corresponding to low-resolution spectra of the source through all the bench. In this figure we can see that output II and III are shifted by approximately one pixel, and scattered on three rows. The optical alignment of the camera bench will be improved to have spots scattered on two rows, without shift. Thus, it will be possible to compare in real time bright and dark fringes for each wavelength.

\begin{figure}
\begin{center}
\begin{tabular}{c}
\includegraphics[height=6.0cm]{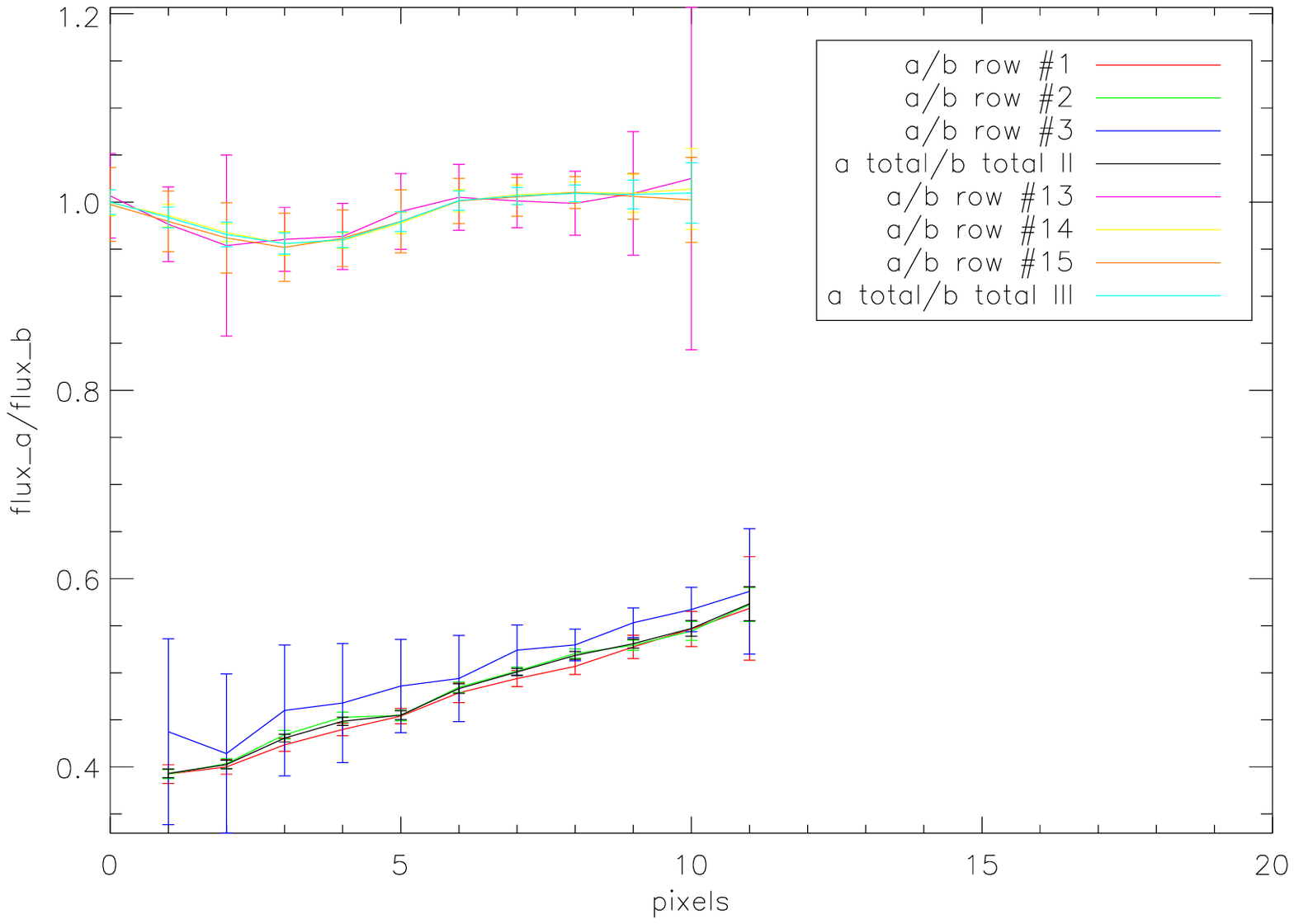}
\includegraphics[height=6.0cm]{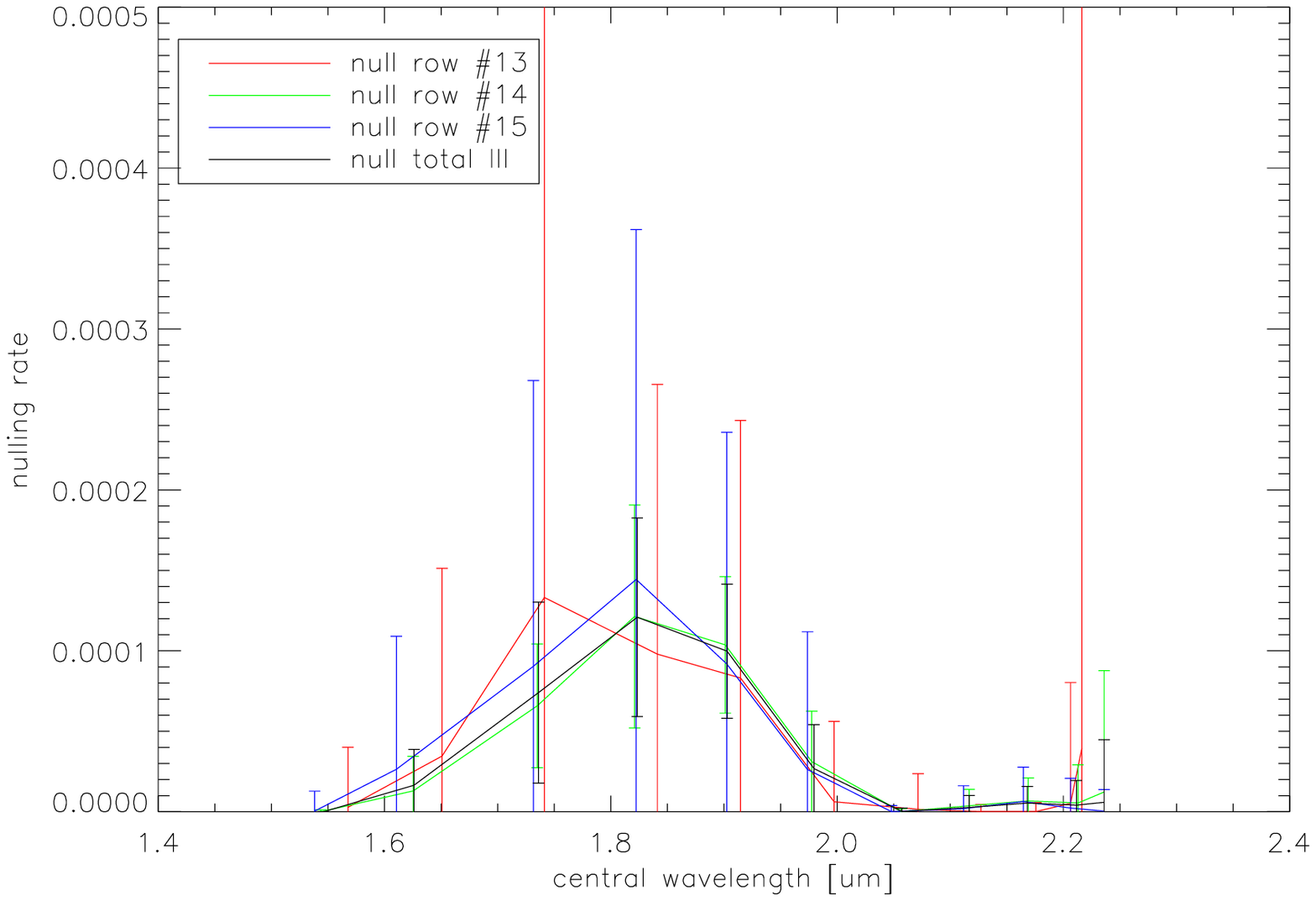}
\end{tabular}
\end{center}
\caption[photom]
{\label{fig:photom}
Ratio between flux in arm a and b (left) and theoretical impact on null depth (right).}
\end{figure}

Figure \ref{fig:spII-III} shows that flux in output III is well balanced, unlike output II. This is consistent with the MMZ design. In figure \ref{fig:photom}, we calculate the ratio between the two arms (left) and the impact of theoretical photometric unbalance on null depth (right). Total photometric unbalance is $1.2\%$. But there is a visible chromatic effect on this value, maybe due to a misalignment of the bench, or injection in single mode fibers. Locally, the unbalance reaches 5\%, with a corresponding nulling of $10^{-4}$ as illustrated on figure \ref{fig:photom} (right). After optimization, the flux unbalance should not be a limitation for our $10^{-4}$ nulling goal. The observed pic at pixel 7-8 (2.13~\mum) should come from a non-linear effect in the optical fiber that doubles the wavelength of the initial 1.064~\mum laser used for the supercontinuum source.

\subsection{Chromatic effects}
\label{sec:chromatic}

\begin{figure}
\begin{center}
\begin{tabular}{c}
\includegraphics[height=6.0cm]{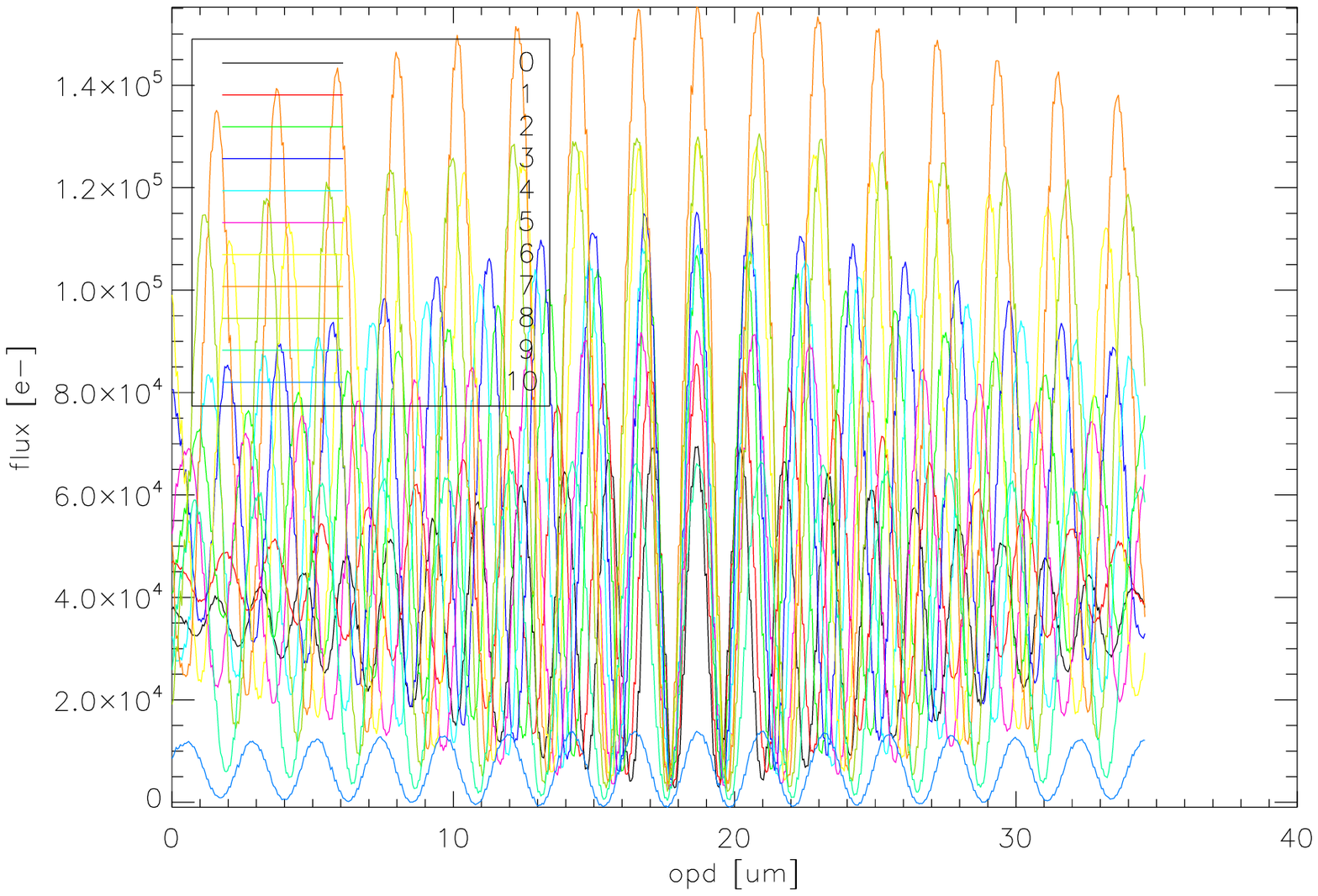}
\includegraphics[height=6.0cm]{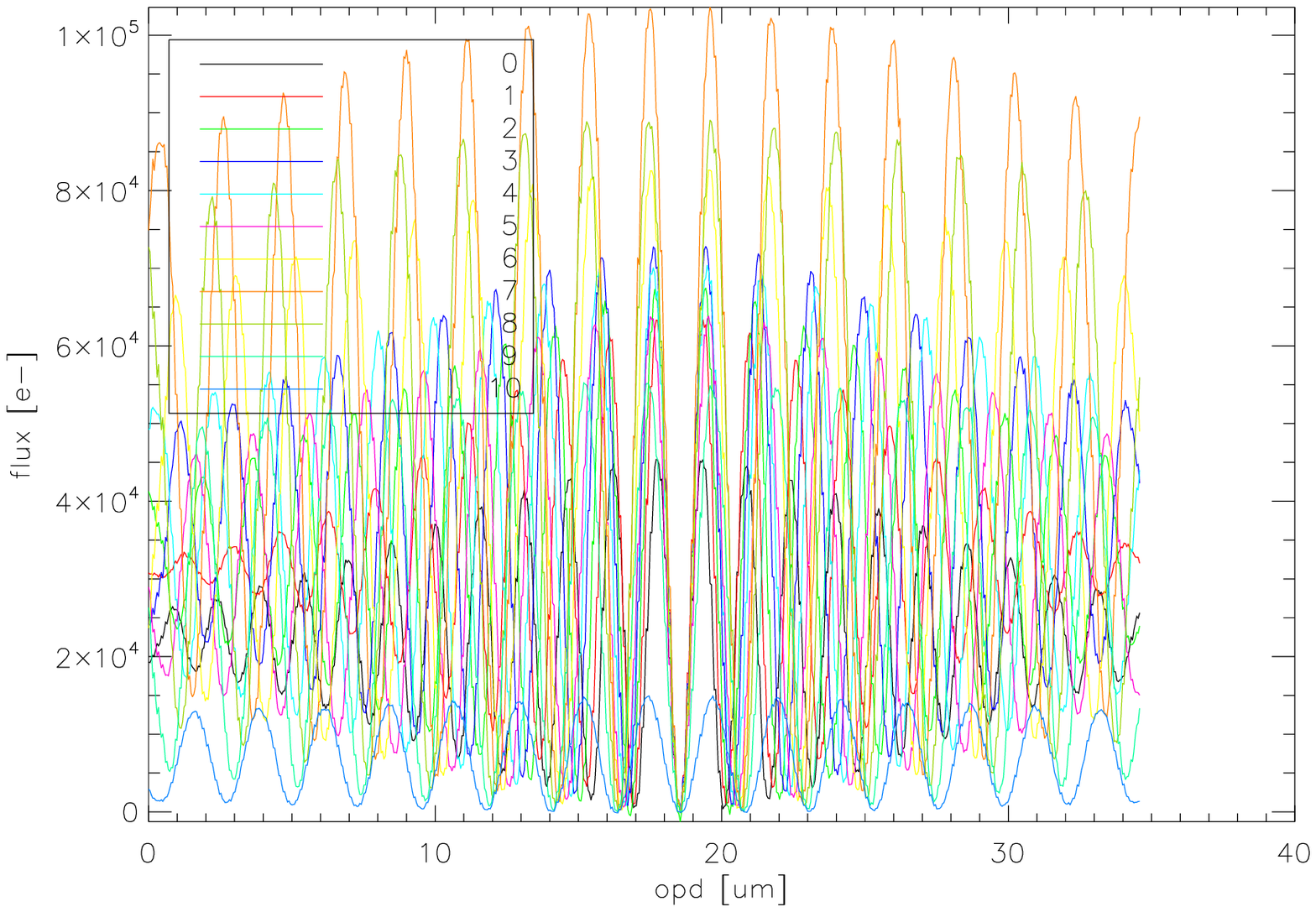}
\end{tabular}
\end{center}
\caption[frII-III]
{\label{fig:frII-III}
Fringes in output II (left) and III (right), obtained by summing three pixels on each column. Thus there are eleven curves corresponding to different wavelengths.}
\end{figure}

Figure \ref{fig:frII-III} is obtained by producing an OPD ramp with the actuating mirrors. We can clearly see that in output II, the achromatic phase is near 0, in contrast to output III, for which achromatic phase is near $\pi$.

\begin{figure}
\begin{center}
\begin{tabular}{c}
\includegraphics[height=6.0cm]{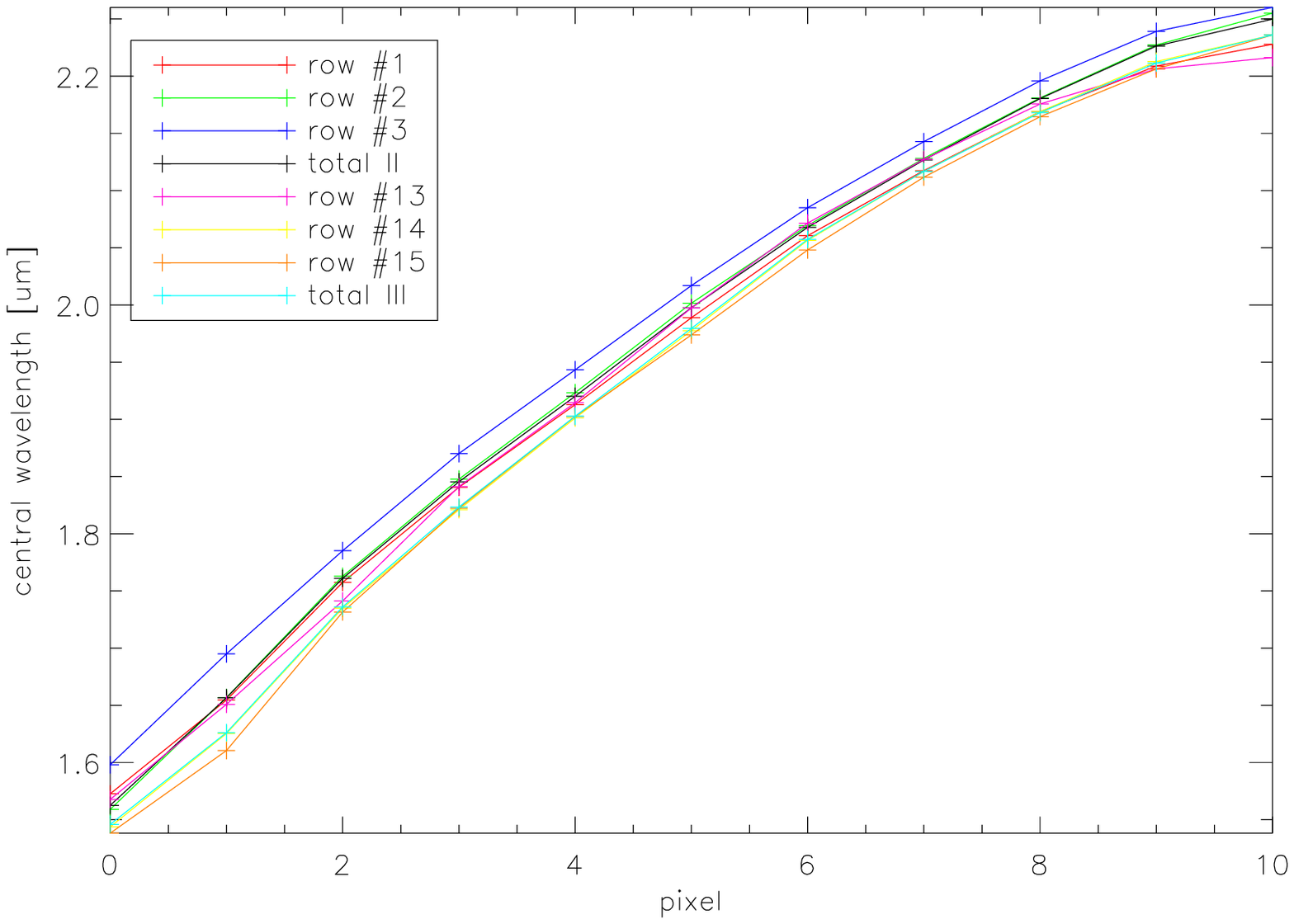}
\includegraphics[height=6.0cm]{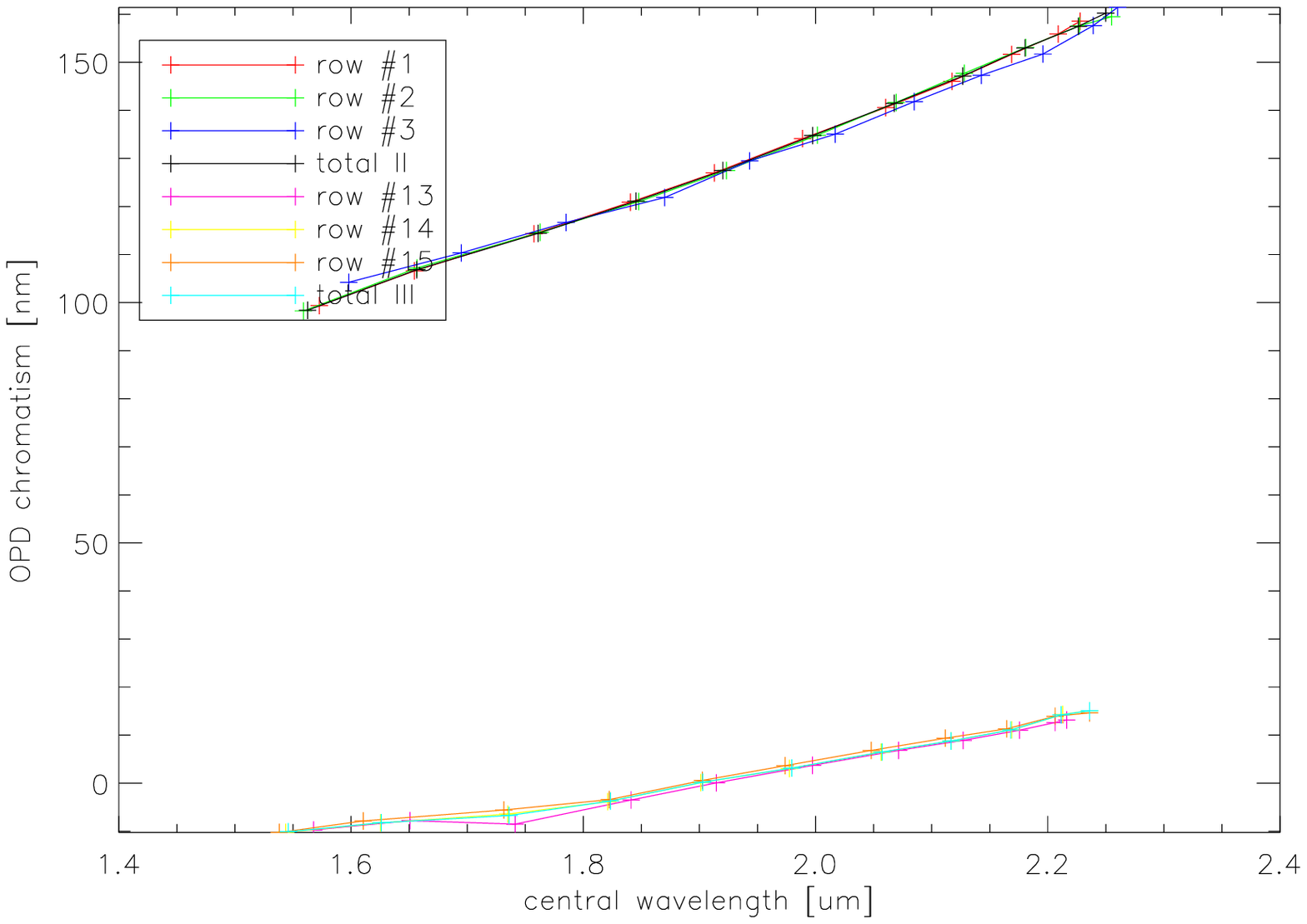}
\end{tabular}
\end{center}
\caption[lambdas]
{\label{fig:lambdas}
Central wavelengths on each pixels, and on each column, for the two outputs (left), and position of fringe minimum for output III and fringe maximum for output II (right). For output III, the ideal case is a constant zero value: all minimums are at the same OPD.}
\end{figure}

\begin{figure}
\begin{center}
\begin{tabular}{c}
\includegraphics[height=6.0cm]{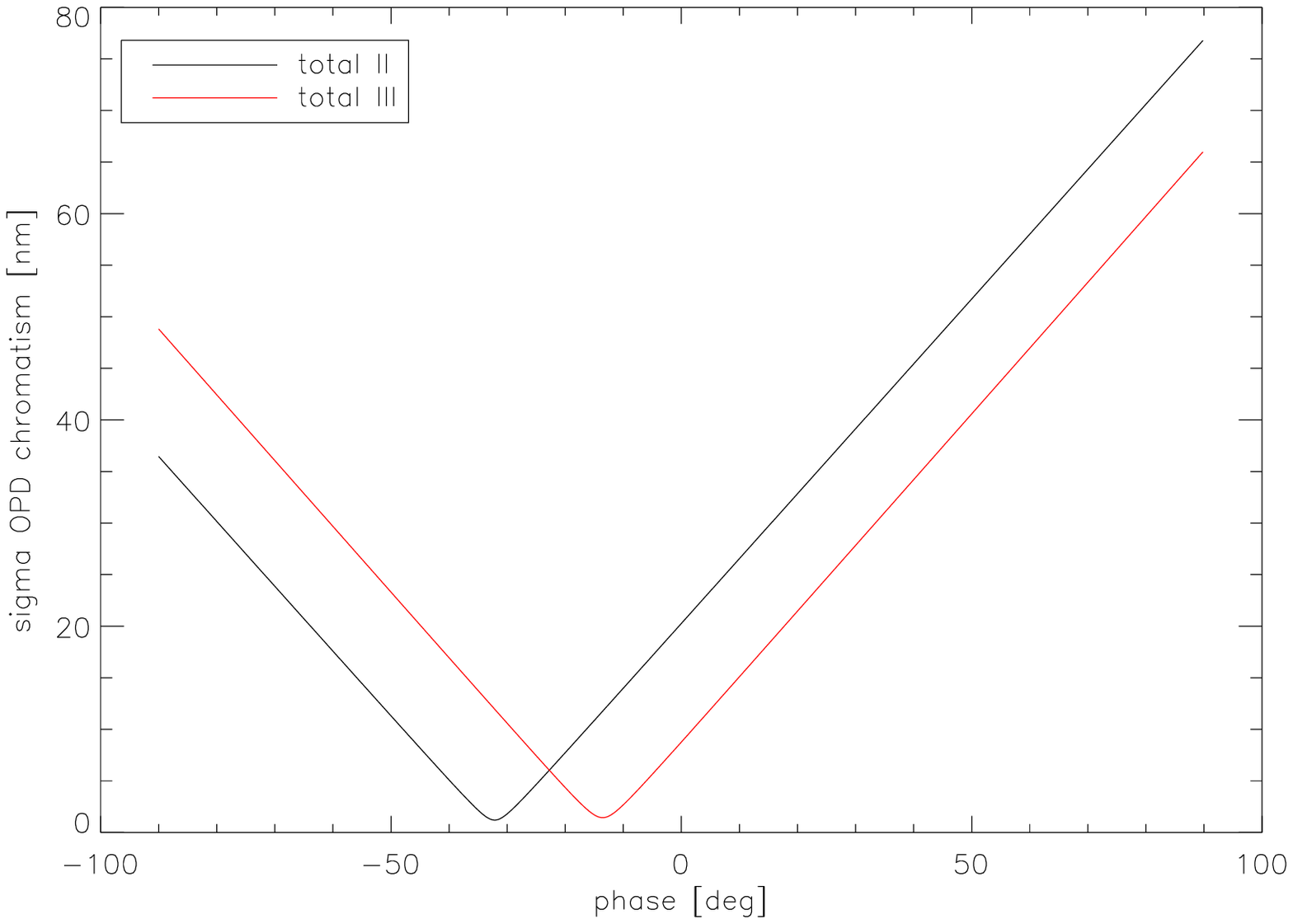}
\includegraphics[height=6.0cm]{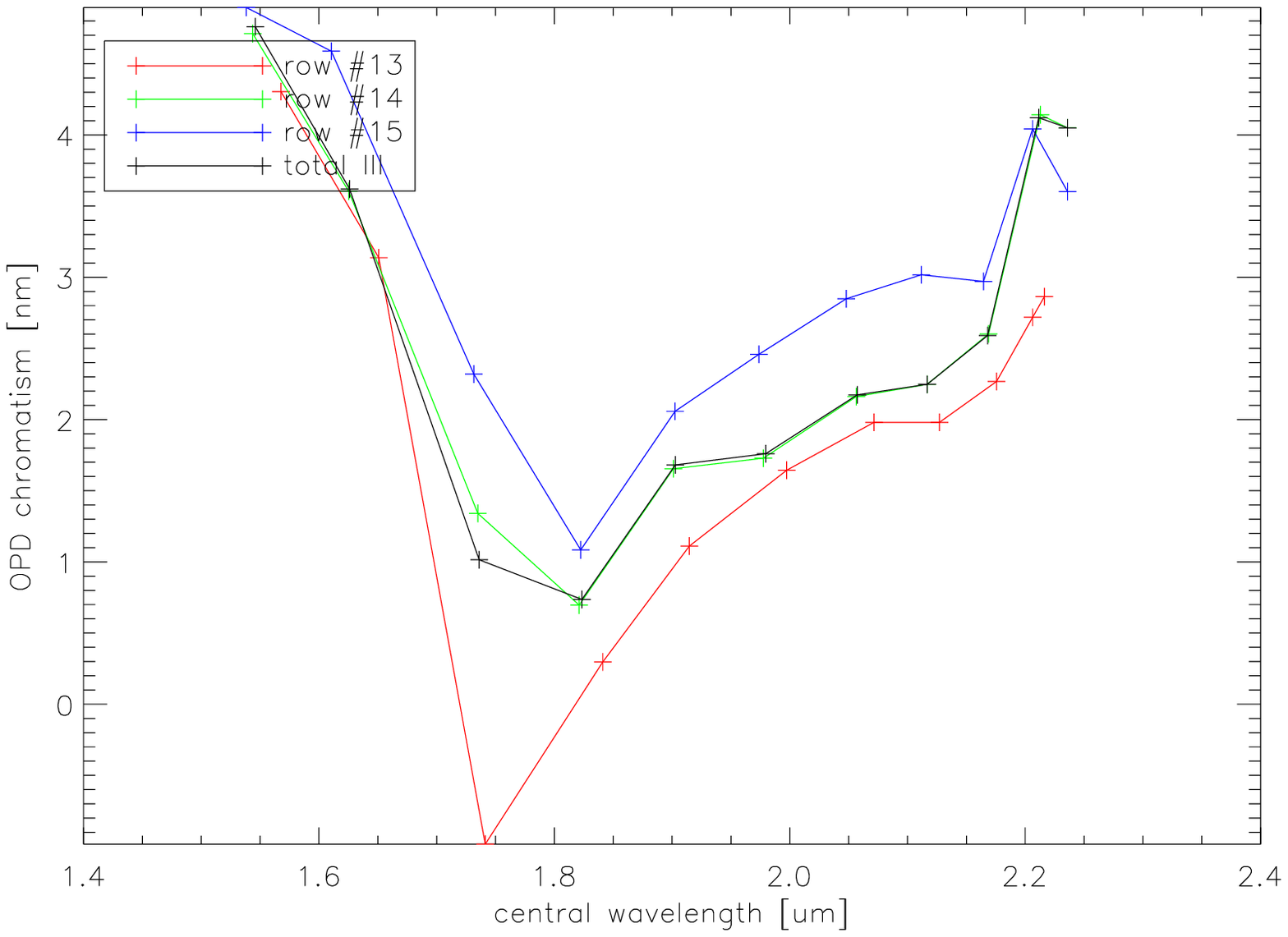}
\end{tabular}
\end{center}
\caption[chrom_min]
{\label{fig:chrom_min}
Standard deviation of OPD versus phase shift considered for the two outputs (left), and minimal chromatic aberration at achromatic phase (right).}
\end{figure}

We can extract a lot of information from these data: By fitting perfect sinusoids to real data on each pixel, we obtained central wavelengths (figure \ref{fig:lambdas} left) on each pixel. As expected, the range of those wavelengths is $\approx[1.6-2.3]\mum$, limited for short wavelengths by the detector and for long wavelengths by the source spectrum. With the fitted sinusoids, we can also determine the chromatic aberration of the two outputs (figure \ref{fig:lambdas} right). Left figure \ref{fig:chrom_min} shows the chromatic shift of each phase compared to the ideal phase (0\degre for output II, 180\degre for output III). So the ideal case would be two line segments passing through the origin. For output II, we were expecting a shift of achromatic phase, because of the unbalanced number of reflexions and transmissions. But for output III, we observe a minimum chromatic error at -13\degre, so achromatic phase is not at 180\degre but 167\degre. The source of this value is currently uncertain, but we assume that it comes from misalignment of beam splitters of the interferometer. Future work will confirm that hypothesis. In any case, we have available a phase shift compensator composed of 4 dispersive plates that can manage this kind of error.

 Right figure \ref{fig:chrom_min} simulates the minimal chromatic effect that will remain after correction by the compensator. This residue is induced by inhomogeneity of coating thickness of beam splitters. Those data are consistent with what has been simulated for the theoretical study of the bench.

\subsection{Polychromatic nulling rate}
\label{sec:poly_null}

We did not perform a real nulling measurement with a stabilized OPD in that configuration, but with the data obtained, we can estimate null depth to a few $10^{-3}$ over the total wavelength range. This value is quite promising for future works, but need to be improved. The chromatic phase shift is the main error we will have to manage, using the phase shift compensator. That should greatly enhance the quality of the null depth.

\section{SOME PRACTICAL AND MORE GENERAL FIRST LESSONS}
\label{sec:LESSONS}

\subsection{Thermal stability}
\label{sec:thermal}

From existing data, we already can infer some general trends and first lessons. By instance, the current MMZ design proved to have a 600~nm/K OPD drift between science and FS channels. In fact, this phenomenon is due to standard mechanics used for supporting the optics and to instability of the mechanism driving the L3a plate. The optics (CaF2) form a nearly auto-compensated optical setup, as the refractive index variations almost compensate the thermoelastic and geometrical effects. An improved MMZ with specially designed optical supports has been considered but could not be afforded in the frame of PERSEE. With sufficient funding, a factor of at least 10 could be saved. This leads to a differential OPD sensitivity of about 50~nm/K. Thus, the thermal stabilization need would be $\pm20$~mK on a box of typically 30x30x10~cm size. The rest of the optical bench could be stabilized at 0.1~K or even 1~K (the driving constraint there is in fact the thermal background in the measurement itself, not the mechanical stability). This seems quite affordable in space, even in non L2 orbits. The situation could even be improved by a clever use of the 4~MMZ outputs which might help to correct the drift due to temperature at regular intervals. To conclude, for an intermediate mission like PEGASE, which is not requiring extreme cooling of the detector or the optics, an earth orbit is not totally ruled out, as first assumed in previous studies. This could be a great simplification as compared to a L2 mission, which remains nevertheless necessary for the final full mission.

\subsection{1~nm cophasing feasibility}
\label{sec:cophasing}

With existing data, we can prove that cophasing two arms on ground at better than 1~nm~rms is feasible on a quite big optical setup in a standard building, but one has to take care to many details. By instance, electronics have to be implemented in an adjacent room to avoid unnecessary noise. The piezoelectric devices can not use their internal strain gauges which introduce too much electrical noise. The stiffness of the mechanical mounts supporting the optics has to be studied with care and a specification of at least 150~Hz for the first mechanical mode seems to us a good rule. Typically, on our bench, the mechanical modes of the optics mounts are located between 80 and 170~Hz and react clearly to acoustic solicitations. The FS design using spatial modulation provides OPD measurements with up to 1~kHz sampling frequency which is extremely useful to implement an adequate compensation of microvibrations for both on ground and in flight perturbations.

Concerning filtering of mechanical perturbators on board a space mission, no experimental results are yet available, but the simulations give good hope that harmonic perturbations can be significantly reduced. This is a good step toward the feasibility of a satellite pointing system using by instance reaction wheels instead of more complex devices.

\subsection{Fine pointing}
\label{sec:pointing}

As far as fine pointing is concerned, a 60~mas stability can be reached at payload level, after the afocal systems. This corresponds to a $1/100^{th}$ of a pixel with a PSF spread over 4 to 5~pixels. It confirms that this point should not be a problem in the future, as expected. By instance, in the PEGASE case, the requirement is 600~mas in the same optical space, a factor of 10 higher.

The use of piezoelectric devices within the payload relaxes the pointing requirements at platform level. The best solution is probably to use a spectral band near $1~\mum$ and collect an annular portion of light in some optic before the combining device, instead of using dichroic plates which would introduce chromatic aberrations and are more difficult to qualify for space use due to coating. The spacecraft requirements will be directly related to the piezoelectric angular stroke (about $\pm100$~arcsec mechanical, $\pm200$~arcsec optical) and the total angular magnification of the system (20 to 40 in the case of PEGASE, about 100 - 150 for DARWIN). In the PEGASE case, we can estimate that the required satellite pointing is about 10~arcsec (residual bias after calibration and stabilization).

\section{CONCLUSION}
\label{sec:CONCLUSION}

Although we did not reach our full objectives yet, we have already a lot of interesting data and very promising results. 1~nm OPD control and 1\% flux unbalance are nearly achieved. The best monochromatic null measured is $3\cdot10^{-5}~\pm~3\cdot10^{-6}$ at 2.32~\mum. The wide band nulling is currently limited by chromatic effects. But this issue will be solved soon by the phase shift compensator. Thus, we are quite confident that PERSEE will be optimized before the end of 2010. Then it will enter an exploitation phase where experiments will be carried out by various teams. This phase will be structured by a scientific group which will collect and sort proposals. At that point, the use of PERSEE will be open to international teams.

\acknowledgments

Julien Lozi's PhD is funded by CNES and ONERA. PERSEE bench is supported by the region \^Ile de France.



\begin{thebibliography}{10}

\bibitem{Leger96}
{L{\'e}ger}, et~al., ``{The DARWIN project},'' {\em
  Astrophysics and Space Science}~{\bf 241},  135-146 (1996).

\bibitem{Leger07}
{L{\'e}ger}, A., et~al., ``{DARWIN mission proposal to ESA},'' {\em ESA's Cosmic
  Vision Call for Proposals}, (2007).

\bibitem{Beichman99}
{Beichman}, C.~A., et~al., ``{The
  Terrestrial Planet Finder (TPF) : a NASA Origins Program to search for
  habitable planets},'' (1999).

\bibitem{Lawson00}
{Lawson}, P.~R., ``{Principles of Long Baseline Stellar Interferometry},'' (2000).

\bibitem{Danchi06}
{Danchi}, W.~C., et~al., ``{Scientific rationale for exoplanet characterization from 3-8 microns: the FKSI mission},'' {\em Proc. SPIE} {\bf 6268}, 626820 (2006).

\bibitem{Ollivier07}
{Ollivier}, M., et~al., ``Pegase, an infrared interferometer to study stellar
  environments and low mass companions around nearby stars,'' {\em ESA's Cosmic
  Vision Call for Proposals}, (2007).

\bibitem{Leduigou06b}
{Le Duigou}, J.~M., ``{Pegase: A Free Flying Interferometer for the
  Spectroscopy of Giant Exo-Planets},'' {\em ESA Special
  Publication} {\bf 621}, (2006).

\bibitem{Cassaing08}
Cassaing, F., et~al., ``{Persee: a nulling demonstrator with real-time correction of external disturbances},'' {\em Proc. SPIE}~{\bf 7013}(1), 70131Z (2008).

\bibitem{Coudeduforesto06}
{Coud{\'e} du Foresto}, V., et~al., ``{ALADDIN: an optimized nulling ground-based demonstrator for DARWIN},'' {\em Proc. SPIE} {\bf 6268}, 626810 (2006).

\bibitem{Houairi08b}
{Houairi}, K., et~al., ``{PERSEE, the dynamic nulling demonstrator: Recent progress on the cophasing system},'' {\em Proc. SPIE}, (2008).

\bibitem{Jacquinod08}
Jacquinod, S., et~al., ``{PERSEE}: description of a new concept for nulling interferometry recombination and opd measurement,'' {\em Proc. SPIE} {\bf 7013}, (2008).

\bibitem{Henault10}
{H\'enault}, F., et~al., ``{Review of OCA activities on nulling testbench PERSEE},'' {\em Proc. SPIE} {\bf 7734}, (2010).

\bibitem{Serabyn01}
{Serabyn}, E., et~al., ``{Fully symmetric nulling beam combiners},'' {\em Appl. Opt.}~{\bf 40}(10),  1668-1671 (2001).

\bibitem{Petit08}
{Petit}, C., et~al., ``{First laboratory validation of vibration filtering with LQG control law for Adaptive Optics},'' {\em Optics Express}~{\bf 16},  87-97 (2008).

\end{thebibliography}

\end{document}